\documentclass[hyper]{JHEP3}

\usepackage{amssymb}
\usepackage{amsfonts}
\usepackage{amsbsy}
\usepackage{amsmath}

\def\Z{\mathbb{Z}}
\def\G{\Gamma}

\title{Four-modulus ``Swiss Cheese" chiral models}

\author{Andr\'es~Collinucci, Maximilian~Kreuzer, Christoph~Mayrhofer and Nils-Ole~Walliser\\

Institute for Theoretical Physics, Vienna University of Technology, \\
Wiedner Hauptstr. 8-10, 1040 Vienna, Austria\\

{\tt andres, kreuzer, cmayrh, walliser  AT hep.itp.tuwien.ac.at }

}

\abstract{We study the `Large Volume Scenario' on explicit, new, compact, four-modulus Calabi-Yau manifolds. We pay special attention to the chirality problem pointed out by Blumenhagen, Moster and Plauschinn. Namely, we thoroughly analyze the possibility of generating neutral, non-perturbative superpotentials from Euclidean D3-branes in the presence of chirally intersecting D7-branes. We find that taking proper account of the Freed-Witten anomaly on non-spin cycles and of the K\"ahler cone conditions imposes severe constraints on the models. Nevertheless, we are able to create setups where the constraints are solved, and up to three moduli are stabilized. 
}

\begin{document}

\section{Introduction}

The `Large Volume Scenario' (LVS), developed in \cite{Balasubramanian:2005zx} is a new strategy for stabilizing the K\"ahler moduli in IIB Calabi-Yau orientifold compactifications. This strategy can be seen as a cousin of the KKLT strategy \cite{Kachru:2003aw}. In both cases, one first stabilizes the axio-dilaton and complex structure moduli by means of the flux induced Gukov-Vafa-Witten superpotential, and then one tries to stabilize the K\"ahler moduli by non-perturbative effects such as E3-branes (Euclidean D3-branes), and gaugino condensation. 
The key difference between these two strategies lies in the fact that the LVS admits non-supersymmetric anti-de Sitter minima, whereby the Calabi-Yau volume is exponentially large w.r.t. the size of the E3-brane, and, at fixed $g_s$, it is independent of the flux superpotential $W_0$. This latter fact implies that this non-perturbative stabilization of the K\"ahler moduli will not mess up the complex structure stabilization. Other advantages of this scenario are explained in \cite{Denef:2008wq}.

The key requirement to construct an LVS model, is to find a Calabi-Yau threefold with $h^{2,1}>h^{1,1}>1$, and such that the volume of the manifold is driven by the volume of a single `large' four-cycle, and that the rest of the four-cycles contribute negatively to the overall volume. This structure has been dubbed the `Swiss cheese' structure. Because it is possible to make cycles small while keeping the CY large, we can have E3-instantons that make large contributions and have a large volume vacuum. These instanton effects now becoming important, actually compete against $\alpha'$ corrections to the K\"ahler potential.
Having these `small', shrinkable cycles also serves another useful purpose. If one places MSSM-like stacks of D7-branes on them, by going to this large volume limit where these are made small, one effectively decouples the gauge theory on the brane from the UV dynamics encoded by the rest of the Calabi-Yau data. In this way, one addresses the comment in \cite{Wijnholt:2008db}, which points out a drawback of generic models: Namely, that making the volume of the CY large will typically force one to scale up the cycles on which branes are wrapped.
\vskip 2mm
In \cite{Blumenhagen:2007sm}, Blumenhagen et al have shown that the standard two-step model building paradigm, where one first stabilizes the closed string moduli and then introduces MSSM-like D7-branes, is too na\"ive. Such D-branes would intersect the E3-branes used in the non-perturbative stabilization, thereby inducing charged zero-modes. In order for the E3 contribution to the superpotential to be non-vanishing, one would then have to turn on vev's for charged superfields, thereby spontaneously breaking the MSSM-like gauge symmetry.
In that article, a solution to the problem is outlined and explicitly worked out for a three-modulus Calabi-Yau, whereby two intersecting stacks of MSSM-like D7-branes are setup so as not to chirally intersect the E3-brane. 
\vskip 2mm
In this paper, we will address the issue of chiral zero-modes while taking even more stringent constraints into account. Namely, we will take into account the fact that the MSSM-like D7-branes are wrapped on non-spin manifolds, thereby inducing half-integral worldvolume fluxes which themselves induce unwanted, charged zero-modes. These fluxes compensate for the open string worldsheet anomaly discovered in \cite{Freed:1999vc}.

We will \emph{not} attempt to construct realistic MSSM configurations. Our goal will be to have setups with two intersecting D7-brane stacks with unitary gauge groups and bifundamental chiral fermions, that accomodate the LVS scenario of moduli stabilization. These setups can then in principle be used to create inflationary models. 
We will see that requiring zero chiral intersections between the E3-brane and the MSSM-like branes, and between the `hidden' D7-branes (that are needed to saturate the negative D7-charge from the O7-planes) and the rest of the branes, will impose heavy restrictions that will rule out some models.

In order to accommodate two MSSM-like D7-stacks and an E3-brane, all on different `small' cycles, we need Calabi-Yau manifolds with at least four moduli, whereby three preferably come from blow-ups. For this purpose, we will scan through the list of CY hypersurfaces of toric fourfolds encoded as four-dimensional polytopes in \cite{Kreuzer:2000xy}. From this list, we will select all four-modulus CY's of which the polytopes have five vertices, which is the minimal amount of possible vertices for four-dimensional polytopes. This will ensure that three of the four moduli will correspond to divisors that originate from blow-ups. We will then proceed to triangulate all relevant polytopes by using a recently enhanced version of the PALP package \cite{Kreuzer:2002uu,wwwCY}. By then feeding the output of PALP into the SINGULAR \cite{SINGULAR} program, we will obtain the triple intersection numbers with which PALP computes the Mori cones of our models. After eliminating models with equivalent triangulations, we end up with four `Swiss cheese' models. 

We will see, however, that not all four-cycles that contribute negatively to the overall CY volume can be shrunk arbitrarily while preserving the large volume limit. This will corroborate the analysis of \cite{Cicoli:2008va} that gives precise conditions for this to be possible. We will also study the topologies of our four-cycles in detail, and will see that not all rigid, `small', cycles with $h^{0,1}=h^{0,2}=0$ are Del Pezzo surfaces, which is a necessary condition for `shrinkability'.

This paper is organized as follows: In section \ref{sec:lvs}, we briefly review some definitions relevant to $\mathcal{N}=1$ IIB orientifold compactifications, we also review the LVS, and we reiterate the chiral zero-mode issue raised in \cite{Blumenhagen:2007sm}. In section \ref{sec:deffreedwitten}, we review how the Freed-Witten anomaly induces half-integral flux when a D-brane wraps a non-spin cycle. In section \ref{sec:zeromodes}, we explain how we count both neutral and charged zero-modes of E3 instantons. Section \ref{sec:firstmodel} contains our first model. Here, we will be very explicit about our strategy. We will present the toric data, explain how we search for and classify `small' divisors, and then move on to model building. This will be done in a three step procedure: First we build `local' models containing MSSM-like branes and E3-branes without canceling the D7 tadpole. Then we pick an orientifold involution and add `hidden' branes appropriately so as not to intersect the visible sector. Finally, we study the K\"ahler moduli stabilization. This model will only be `half' successful, in the sense that we will be able to solve the chiral intersection problem but will not find a large volume minimum. In section \ref{sec:T4model} we present our second model, which will be successful in this sense, although one unstabilized modulus will remain. In appendix \ref{app:branes} we present the relevant definitions for B-branes such as induced charges, orientifolding, D-terms and constructions of involution invariant D7-branes. Finally, in appendices \ref{sec:T1model} and \ref{sec:nomatter} we present our two remaining models.

All of our results are summarized in table \ref{tab:finalresults}, in the conclusions.

\section{Large volume scenario} \label{sec:lvs}
\subsection{General idea}
We briefly review some definitions for $\mathcal{N}=1$ flux compactifications of type IIB in order to set our conventions. The full superpotential for IIB compactified on a CY threefold $X$ is given by\\
\begin{equation}
 W=\int_X G_3\wedge\Omega_3+ \sum_i A_i\left(S,U\right) e^{- a_i T_i }\,, 
 \label{superpotential}
\end{equation}
where the first term is the GVW potential \cite{Gukov:1999ya}, which stabilizes the complex structure moduli and the axion-dilaton field, $S=e^{-\phi}+iC_0$. The second term takes into account non-perturbative corrections to the superpotential. We focus here on corrections due to the presence of E3-brane instantons, in which case the functions $A_i$ only depend on the axion-dilaton and the complex structure moduli.\footnote{In general non-perturbative corrections can also arise from gaugino condensation from wrapped D7-branes.} Here $\tau_i$ denotes the volume of the divisor $D_i$, and $\rho_i$ is the corresponding axion field originating from the RR four-form:
\begin{equation}
\tau_i= \frac{1}{2}\int_{D_i} J\wedge J=\frac{1}{2} \kappa_{ijk} \,t^j\,t^k\,, \quad {\rm and} \quad \rho_i = \int_{D_i} C_4\,.
\end{equation}
The $\kappa_{ijk}$ coefficients determine the triple intersection numbers given a basis of integral two-forms $\{\eta_i\}\in H^{1,1}\left(X,\mathbb{Z}\right)$, in which we will choose to expand the K\"ahler form:
\begin{equation}
J = \sum_i t_i\,\eta_i\,.
\end{equation} 

The K\"ahler potential with its leading $\alpha'$-correction \cite{Becker:2002nn} takes the following form 
\begin{eqnarray}
 K &=& -2 \ln\left({\hat{\cal V}} + \frac{\xi}{2 g_s^{3/2}}\right)- \ln\left(S+\bar S \right)-\ln \left(-i\int_X \Omega\wedge\bar\Omega\right)\,.
 \label{kaehlerpotgeneral}
\end{eqnarray}
Where $\xi=-\frac{\zeta\left(3\right)\chi\left(X\right)}{16 \pi^3}$ encodes the perturbative $\alpha'$-correction in terms of the Euler characteristic of $X$. The symbol ${\hat{\mathcal{V}}}$ denotes the volume of the CY in the Einstein-frame, where the metric is expressed in terms of the string-frame metric by $g_{\mu\nu,E}=e^{-\phi/2}g_{\mu\nu}$. The volume in the string-frame is given by
\begin{equation}
	\mathcal{V}=\frac{1}{3!}\int_X J\wedge J\wedge J=\frac{1}{6} \kappa_{ijk}\,t^i\,t^j\,t^k\,.
\end{equation}
Note that in computing the volume of the CY we assume that NSNS fluxes have stabilized the background value of the dilaton. Hence, we may effectively treat the latter as a constant, and readily switch frames. Strictly speaking, this is only a large volume approximation, as the dilaton will vary strongly in the vicinity of the D7-brane and the O7-plane. Since we will mainly work in the string-frame, from now on we will explicitly denote quantities in the Einstein-frame by a hat symbol.

The four-dimensional scalar potential for all moduli fields gets contributions from both F- and D-term potentials. The F-term has the following form
\begin{equation}\label{equ:f-term-general}
V_F = e^K \left( \sum_{i=T,S,U} K^{i\bar j}D_i W D_{\bar j} \bar W -3 |W|^2 \right)\,,
\end{equation}
where the sum runs respectively over K\"ahler structure, the axion-dilaton and the complex structure moduli. The non-perturbative term in the superpotential depends explicitly on the K\"ahler moduli, $T_i=e^{-\phi}\tau_i+i\rho_i$, and thus breaks the no-scale structure of the superpotential. 

We are interested in CY manifolds characterized by a volume function of the following shape
\begin{equation}
\mathcal{V}\sim \tau_l^\frac{3}{2}-\sum_{s=1}^{h^{1,1}-1}\tau_s^\frac{3}{2}\,.
\end{equation}
The important property of this function lies in the fact that there is one four-cycle that contributes positively to the volume, and the remaining three contribute negatively. This means that, in principle, one can take a limit where the positively contributing cycle is taken large, and the other three are sent small, while keeping the overall volume of the CY large. Hence, the cycle with volume $\tau_l$ will be referred to as a `large' cycle, and the remaining ones as `small' cycles. For this reason these manifolds are colloquially referred to as `Swiss cheese' CY manifolds. 

The reason why one would like to have such a CY is that it allows for the LVS \cite{Balasubramanian:2005zx}, which we will now briefly describe.
Inserting (\ref{superpotential}) and (\ref{kaehlerpotgeneral}) in the above formula for the F-term \eqref{equ:f-term-general}, the potential has three parts: Two non-perturbative terms depend explicitly on the K\"ahler moduli, and one term accounts for the $\alpha'$-corrections: $V_F=V_{np1}+V_{np2}+V_{\alpha'}$. In the large volume regime these terms behave like
\begin{eqnarray}
 V_{np1} & \sim & \frac{1}{\mathcal{\hat{V}}} a_s^2 \vert A_s\vert^2\, \left(-\kappa_{ssj} \,t^j\right)e^{-2 a_s \hat{\tau}_s}e^{K_{cs}}+\mathcal{O}\left(\frac{e^{-2a_s\hat{\tau}_s}}{\mathcal{\hat{V}}^2}\right) \, \\
 V_{np2} & \sim & -\frac{a_s\hat{\tau}_se^{-a_s\hat{\tau}_s}}{\mathcal{\hat{V}}^2}\vert A_s W_0\vert \,e^{K_{cs}} +\mathcal{O} \left(\frac{e^{-a_s\hat{\tau}_s}}{\mathcal{\hat{V}}^3}\right)\,  \\
 V_{\alpha'} & \sim &\frac{3 \hat{\xi}}{16 \mathcal{\hat{V}}^3}\vert W_0 \vert^2 \,e^{K_{cs}} +\mathcal{O}\left(\frac{1}{\mathcal{\hat{V}}^4}\right)\,.
\end{eqnarray}
$V_{np1}$ is positive and proportional to self-intersection of the small cycle. Since we require $h^{2,1}>h^{1,1}$, also $V_{\alpha'}$ contributes positively to the potential. The second term, instead, contributes negatively.  
If we consider the decompactification limit, maintaining $a_s \hat{\tau}_s=\ln \mathcal{\hat{V}}$, the three terms become proportional to the inverse third power of the CY volume, thus they are all on equal footing. At this point, the potential is negative. But for increasing $\hat{\mathcal{V}}$, $V_{np2}$ grows faster than $V_{np1}+V_{\alpha'}$. Due to the positive contribution of $V_{np1}$ and $V_{\alpha'}$ the potential starts positive by small volume values, then reaches a negative minimum and afterwards approaches zero from below for asymptotically large values of the volume. This assures the existence of a local anti-de Sitter minimum at finite volume. The `Swiss cheese' shape of the manifold is needed here to keep the cycle $\hat{\tau_s}$ logarithmically small compared to the overall volume.

Assuming that the complex structure moduli and the axio-dilaton have been stabilized via the GVW superpotential, we can rewrite the F-term potential for the K\"ahler moduli in the large volume limit following \cite{Bobkov:2004cy},\cite{Cicoli:2008va},\cite{Cicoli:2007xp} and \cite{Blumenhagen:2007sm}:
\begin{eqnarray}
V_F &=& \frac{1}{\hat{\cal V}^2}\Big(-4 \pi^2 {\rm Vol}\left(D_{E3}\cap D_{E3}\right)\hat{\cal V} \left|A_{E3}\right|^2 e^{-4\pi\hat\tau_{E3}}\nonumber\\
& &-4\pi \hat\tau_{E3} e^{-2\pi \hat\tau_{E3}} \left|A_{E3} W_0\right|+\frac{3}{4}\frac{\hat\xi}{\hat{\cal V}}\left| W_0\right|^2\Big)\,. \label{equ:f-term}
\end{eqnarray}

Let us now discuss D-terms. The several D7-branes wrapped on divisors $D_i$ give rise to the following D-term:
\begin{equation}\label{equ:d-term-general}
 V_D = \sum _{i=1}^N \frac{1}{\Re (f_i)}\left(\sum_j Q_j^{(i)}|\phi_j|^2 -\hat{\xi}_i\right)^2\,,
\end{equation}
where the $\phi_i$ are chiral fields charged under the gauge symmetries of the D7-branes.
This potential is determined by the real part of the gauge kinetic functions 
\begin{equation}
 \Re (f_i) = e^{-\phi} \frac{1}{2}\int_{D_i} J\wedge J- e^{-\phi}\int_{D_i} {\rm ch}_2 \left(\mathcal{L}_i-B\right)=\hat\tau_i-e^{-\phi} c_i
\end{equation} 
and the Fayet-Iliopoulos terms
\begin{equation}
\hat \xi_i =-\Im\left(\frac{1}{\hat {\cal V}}\int_X e^{-(B+i\,\hat{J})}\,\Gamma_i\right)  
\end{equation}
Here $\Gamma_i$ denote the charge vectors of the D7-branes. See appendix \ref{sec:dbranecharges&dsz} for a definition thereof.

\subsection{Incorporation of D7-brane stacks}
In order to combine the closed string moduli stabilization with a string theoretic realization of the MSSM, the standard paradigm in IIB string theory describes the gauge groups as arising from D-brane stacks, and the chiral matter from intersections between stacks. This necessarily requires the incorporation of D7-branes, and therefore O7-planes.

As was explained in \cite{Blumenhagen:2007sm}, the standard strategy of first stabilizing all closed string moduli and then adding MSSM-like D7-brane stacks has a serious pitfall. As we will explain in the next section, the D7-branes will in general be forcefully magnetized. Since they will generically intersect the E3-branes, the E3-D7 strings will correspond to chiral zero-modes of the instanton that are charged under the MSSM-like gauge groups. Therefore, in order to saturate the instanton path integral, any non-zero contribution will have to be accompanied by a multiplicative factor of charged superfields.
Since we want our LVS models to serve as a first step in creating models that describe the inflationary epoch, during which energies were above the electro-weak breaking scale, we want to keep the MSSM-like gauge group unbroken. This means that charged superfields must have zero vev's, which will then force such charged superpotentials to vanish. 

The strategy is then to engineer our models as follows: We will have one E3-brane placed on a `small' four-cycle, and two MSSM-like D7-branes with unitary gauge groups placed on the two remaining four-cycles. Finally, to cancel the total D7 tadpole by the O7-plane, we need a `hidden' D7-brane. We will impose the following constraints on the chiral intersections between the branes:
\begin{enumerate}
\item Both MSSM-like D7-branes have no net chiral zero-modes with the E3-brane, but do have chiral matter amongst themselves.
\item The `hidden' D7-brane has no chiral intersections with either the MSSM-like branes, nor the E3-brane.
\end{enumerate}

In the next section, we explain more clearly why D7-branes are forcefully magnetized.
\section{Freed-Witten anomaly} \label{sec:deffreedwitten}
In order for the open string worldsheet theory to be consistent, the submanifold on which a D-brane is wrapped must be chosen with care. In \cite{Freed:1999vc}, Freed and Witten worked out two types of pathologies that can arise. 

If a D-brane is wrapped on a submanifold $W$, such that the pullback of the NSNS three-form field-strength onto $W$ is non-trivial, i.e. $\imath^*(H) \neq dB$, then the open worldsheet theory has a fatal anomaly that can only be compensated by having lower brane worldvolumes end on $W$. We will not thoroughly analyze this issue in this paper, but will make remarks about it whenever possible. 

The other possible pathology has to do with the topology of the submanifold $W$ itself. If $W$ does not admit a spin structure, this leads to a worldsheet anomaly, unless one compensates this by `twisting' the would-be spin bundle with a would-be U(1)-bundle (see \cite{Evslin:2006cj} for a pedagogical explanation of this). Pragmatically, this means that one has to turn on a `half-integral' Born-Infeld flux equal to $F = -c_1(N_W)/2$. In general, the total flux on a D-brane will be of the form
\begin{equation}
F = -\frac{c_1(N_W)}{2}+\Delta F\,,
\end{equation}
where $\Delta F \in H^2(W, \Z)$. Although this half-integral shift is in some sense artificial, it must be taken seriously for all practical purposes: It will induce lower brane charges and will contribute to the chiral intersections \eqref{dsz}, as explained in \cite{Minasian:1997mm}. This latter fact severely constrains the possibility of generating neutral superpotentials, i.e. superpotentials arising from E3 instantons that have no chiral intersections with the D7-branes in the setup.\\

Although we will not in general be able to determine, whether or not a submanifold is spin, we will at least be able to test, whether the `visible' effect of the half-integer shift can be canceled by a \emph{bona fide} integral flux in $H^2(W,\Z)$. More precisely, we will establish a necessary criterion to test for this possibility.

The formula \eqref{dsz} for the chiral intersection between two branes depends on the Born-Infeld fluxes only through the charges they induce. For branes wrapping four dimensional submanifolds, the Dirac-Schwinger-Zwanziger (DSZ) product depends on the D7 and induced D5 charges seen in the total charge vector \eqref{defsmoothd7charges}. Therefore, the only way the half-integer shift in the flux can do harm, is through the part that survives the push-forward operation
\begin{equation}
F \mapsto \Big(\int_W F\cdot \imath^*(D_A) \Big)\,\tilde{D}^A\,.
\end{equation}
Now suppose there is a two-form $\gamma \in H^2(W, \Z)$ such that 
\begin{equation}
\int_W \big(-\frac{c_1(N_W)}{2}+\gamma)\cdot \imath^*(D_A) = 0 \quad
\forall \quad D_A \in H^2(X, \Z)\,.
\end{equation}
Clearly, $\gamma$ cannot be a pulled-back from $\gamma \neq \imath^*()$, since it would have to emanate from a half-integral form in $X$. It could, however, be a two-form that can be decomposed into a pulled-back part in $H^2(X, \mathbb{Q})$ and a part orthogonal to this. Such forms are referred to as `gluing vectors', (see \cite{Denef:2007vg} for definitions). 
Be that as it may, the two-form $\gamma$, which is assumed to be of type $(1,1)$, must be Poincar\'e dual to some linear combination of holomorphic curves on $W$
\begin{equation}
[\gamma] = \sum_i n_i [C_i]\,, \quad [C_i] \in H^2(W, \Z)\,, \quad n_i \in \mathbb{Z}\,.
\end{equation}
By virtue of the fact that $W$ is holomorphically embedded in $X$, and that these curves are holomorphically embedded in $W$, the latter are also holomorphically embedded in $X$. Therefore, the induced D5-charges can be written as follows:
\begin{eqnarray}
q_{D5, A} \equiv \int_W \gamma \cdot \imath^*(D_A) &=& \sum_i n_i\, \int_{C_i} \imath^*(D_A)\,, \\
&=& \sum_i n_i\, \int_{\imath_*(C_i)} D_A\,,
\end{eqnarray}
where, in the last line, we integrate the $D_A$ over the push-forwards $\imath^*(C_i)$ of the curves. Since the latter are well-defined classes in $H_2(X, \Z)$, these D5-charges must be integers.\\

\noindent In conclusion, we will apply the following rule:
\vskip 2mm
\noindent A D7-brane wrapped on a divisor $W$ will at least carry a half-integer flux $F = -c_1(N_W)/2$. If the induced D5-charges are not all integers, then this half-integral shift cannot be compensated by turning on a \emph{non pulled-back} flux. If they are all integers, then more information is needed to decide.

\section{Instanton zero-mode counting} \label{sec:zeromodes}
The spacetime effects of D-instantons and their zero-modes can be described by means of CFT. Detailed accounts of this topic in the IIA setting can be found in \cite{Akerblom:2007nh, Blumenhagen:2006xt, akerblomphd}, and in the IIB setting in \cite{Kachru:2008wt}.

\subsection{Neutral zero-modes}
We are interested in finding E3-branes that will induce a four-dimensional non-perturbative superpotential depending on the complexified K\"ahler modulus corresponding to the divisor of the E3-brane. Witten's well-known criterion for determining whether a specific E3-brane may or may not contribute requires finding an explicit F-theory lift of the IIB setup. One can also work directly in IIB and count the number and type of fermionic zero-modes associated with the E3-brane.

In order for the E3-brane to generate a superpotential as opposed to a higher F-term or a D-term, it cannot have more than two fermionic neutral zero-modes. Neutral zero-modes arise from strings with both end points on the E3. These can be classified into the following three categories:
\begin{enumerate}
\item Universal zero-modes: These strings correspond to the four real scalar fields on the worldvolume theory of the E3 parametrizing transverse motion in four-dimensional spacetime, and their four fermionic superpartners. These modes are model-independent, as the name suggests. 

The integration over these modes must be saturated by operator insertions that will destroy the superpotential structure of the instanton contribution, thereby turning it into a D-term. There are several known mechanisms to get around this issue. One of them is to let the orientifold projection get rid of half of these fermionic zero-modes \cite{Argurio:2007qk, Argurio:2007vqa, Bianchi:2007wy, Ibanez:2007rs}. This requires the E3-brane to be transversally invariant under the orientifold involution, i.e. that it be mapped to itself as a set. Other mechanisms are known (e.g. \cite{Petersson:2007sc}), but we will focus on the orientifold mechanism.

\item Internal motion of the E3: The divisor $D$ on which the E3 is wrapped can have moduli, which will also correspond to scalar fields in the worldvolume field theory. The number of these moduli is given by the number of non-trivial sections of the normal bundle of $D$
\begin{equation}
\# \ {\rm sections} \ =H^0 (D, ND)\,.
\end{equation}
By Serre duality, (or very roughly, by contracting with the holomorphic three-form of $X$), this dimension is equal to the the Hodge number $h^{0,2}(D)$:
\begin{equation}
{\rm dim}\, H^2(D, \mathcal{O}) = h^{0,2}(D)\,.
\end{equation}

\item Wilson lines: If the divisor $D$ has $h^{0,1} \neq 0$, then the worldvolume gauge theory has Wilson line moduli. These can be counted as follows: First, we compute $h^{0,2}$ by counting the number of non-trivial sections of the normal bundle of $D$. Then, we compute the holomorphic Euler characteristic $\chi_h$ of $D$
\begin{equation}
\chi_h = \frac{1}{12}\,\int_X \big(2\,D^3+c_2(X) \cdot D \big)\,. \label{holomorphiceuler}
\end{equation}
From this, we can deduce $h^{0,1} = 1+h^{0,2}-\chi_h$\,.

\end{enumerate}

Imposing $h^{0,2} = h^{0,1} = 0$ is a sufficient criterium for the instanton to contribute to a superpotential. However, the latter may be a charged superpotential, as we will see next.

\subsection{Charged zero-modes}
In \cite{Blumenhagen:2007sm}, a very important issue has been raised concerning the generation of an \emph{uncharged} superpotential. If an E3-brane intersects a D7-brane, the strings stretched between them give rise to bifundamental zero-modes that also need to be soaked up. This requires inserting charged chiral superfields $\Phi_i$ in the path integral, thereby spoiling the generation of an uncharged superpotential, and leading to something of the form
\begin{equation}
W \sim \prod_i \Phi_i \, e^{-T}\,.
\end{equation}
In order for such a term to be non-zero, one must then require that the superfields have vev's. The main point of \cite{Blumenhagen:2007sm} is that one does not want to break the MSSM-like gauge group at the high energy scale of this setup. Hence, in order to generate phenomenologically viable (uncharged) superpotentials, we must require that the E3 does not intersect any other brane present:
\begin{equation}
\langle \G_{E3}, \G_{D7} \rangle = 0\,.
\end{equation}
Searching for setups that satisfy this equation will be the main concern of this paper. The fact that D-branes generically have a half-integral flux that cannot be turned off, as explained in the previous section, will severely restrict the possibility of having setups with several D7-brane stacks with none of them intersecting the E3-branes.

\section{First model} \label{sec:firstmodel}

For the sake of clarity, we will give a very detailed account of this first model. We will be more concise in the subsequent models. For a brief introduction to the geometrical methods we used, see \cite{Kreuzer:2006ax,Denef:2008wq} and the references therein.

\subsection{The resolved $\mathbb{P}^4_{1,2,2,10,15}(30)$ geometry}
\subsubsection{Toric data}
Our first model will be the degree $30$ hypersurface of the weighted projective space $\mathbb{P}^4_{1,2,2,10,15}$. Smoothing out this model requires three toric blow-ups, thereby endowing the CY manifold with four K\"ahler moduli. The following table shows the homogeneous coordinates of the ambient fourfold and their projective weights under the four $\mathbb{C}^*$ actions. 
\begin{table}[h]
\begin{centering}
\begin{tabular}{|c|c|c|c|c|c|c|c||c|}
\hline 
$x_{1}$ & $x_{2}$ & $x_{3}$ & $x_{4}$ & $x_{5}$ & $x_{6}$ & $x_{7}$ & $x_{8}$ & \emph{p}\tabularnewline
\hline
\hline 
15 & 10 & 2 & 2 & 1 & 0 & 0 & 0 & 30\tabularnewline
\hline 
9 & 6 & 1 & 1 & 0 & 1 & 0 & 0 & 18\tabularnewline
\hline 
7 & 5 & 1 & 1 & 0 & 0 & 1 & 0 & 15\tabularnewline
\hline 
3 & 2 & 0 & 0 & 0 & 0 & 0 & 1 & 6\tabularnewline
\hline
\end{tabular}
\par
\end{centering}
\caption{Projective weights under the toric $\mathbb{C}^*$ actions for the resolved $\mathbb{P}^4_{15,10,2,2,1}(30)$ space. The peculiar order of the coordinates is due to PALP's internal computational optimization.}
\label{chargesfirstmodel}
\end{table} 
For the unique triangulation the Stanley-Reisner ideal reads
\begin{equation}
SR=\{x_{1}\,x_{5}, x_{5}\,x_{8}, x_{7}\,x_{8}, x_{1}\,x_{2}\,x_{6}, x_{1}\,x_{2}\,x_{8}, x_{3}\,x_{4}\,x_{5}, x_{3}\,x_{4}\,x_{6}, x_{3}\,x_{4}\,x_{7}, x_{2}\,x_{6}\,x_{7}\}. \label{srfirstmodel}
\end{equation} 
In this notation, the entries are coordinates that are not allowed to vanish simultaneously. For instance, the last entry means that $x_2, x_6$ and $x_7$ cannot vanish simultaneously.
The triple intersection numbers of divisor classes\footnote{Throughout this paper, we will use the sloppy notation where $\eta_i$ can denote a two-form, a second cohomology class, a divisor, and a line bundle whose first Chern class is given by the denoted two-form. It should, however, always be clear from the context which interpretation is appropriate.} in the basis $\eta_{1}=D_{5}$, $\eta_{2}=D_{6}$,
$\eta_{3}=D_{7}$, $\eta_{4}=D_{8}$ are encoded in the following polynomial
\begin{eqnarray}
I_{3} & = & 8\eta_{1}^{3}+8\eta_{2}^{3}-96\eta_{3}^{3}+9\eta_{4}^{3}+3\eta_{1}^{2}\eta_{2}-21\eta_{1}^{2}\eta_{3} \nonumber \\
 &  & -5\eta_{1}\eta_{2}^{2}+\eta_{2}^{2}\eta_{4}-3\eta_{2}\eta_{4}^{2}+45\eta_{1}\eta_{3}^{2}\,.
 \label{intfirstmodel}
 \end{eqnarray}
The K\"ahler form in the basis $\left\{ \eta_{1},\eta_{2},\eta_{3},\eta_{4}\right\} $
is given by
\begin{equation}
J=t_{1}\eta_{1}+t_{2}\,\eta_{2}+t_{3}\,\eta_{3}+t_{4}\eta_{4}.
\end{equation}
The volumes of the corresponding divisors are
\begin{eqnarray}
\tau_1 &=& \frac{1}{10}\, \Big( (15\,t_3 - 7\,t_1)^2 - (3\, t_2-5\, t_2)^2 \Big)\,, \nonumber \\
\tau_2 &=& \frac{1}{6}\, \Big( (3\, t_1-5\, t_2)^2 - (t_2-3\, t_4)^2 \Big)\,, \\
\tau_3 &=& \frac{1}{14} \Big( 3\, t_3^2-3\, (15\,t_3 - 7\,t_1)^2 \Big)\,, \nonumber \\
\tau_4 &=& \frac{1}{2}\,(t_2-3 t_4)^2\,, \nonumber 
\end{eqnarray}
and the volume of the CY manifold is given by
\begin{eqnarray}
\mathcal{V} & = & \frac{1}{630}\left[45t_{3}^{3}-3\left(15\,t_3 - 7\,t_1\right)^{3}-7\left(3t_{1}-5t_{2}\right)^{3}-35\left(t_{2}-3t_{4}\right)^{3}\right] \label{volfirstmodel} \\
 & = & \frac{\sqrt{2}}{3}\left[\frac{1}{7\sqrt{3}}\left(15\tau_{1}+9\tau_{2}+7\tau_{3}+3\tau_{4}\right)^{\frac{3}{2}}-\frac{1}{35}\left(5\tau_{1}+3\tau_{2}+\tau_{4}\right)^{\frac{3}{2}}-\frac{1}{15}\left(3\tau_{2}+\tau_{4}\right)^{\frac{3}{2}}-\frac{1}{3}\tau_{4}^{\frac{3}{2}}\right]. \nonumber
\end{eqnarray} 
It has the expected Swiss cheese form. From this volume formula we
deduce the diagonal basis to be
\begin{eqnarray}
D_{a} & = & 15\eta_{1}+9\eta_{2}+7\eta_{3}+3\eta_{4} \,,\nonumber\\
D_{b} & = & 5\eta_{1}+3\eta_{2}+\eta_{4} \,, \\
D_{c} & = & 3\eta_{2}+\eta_{4} \,, \nonumber \\
D_{d} & = & \eta_{4} \,.\nonumber
\label{diagfirstmodel}
\end{eqnarray} 

In this basis the total volume reads\[
\mathcal{V}=\frac{\sqrt{2}}{3}\left(\frac{1}{7\sqrt{3}}\tau_{a}^{\frac{3}{2}}-\frac{1}{35}\tau_{b}^{\frac{3}{2}}-\frac{1}{15}\tau_{c}^{\frac{3}{2}}-\frac{1}{3}\tau_{d}^{\frac{3}{2}}\right),\]
and the triple intersections can be rewritten as
\begin{equation}
I_{3}=147\,D_{a}^{3}+1225\,D_{b}^{3}+225\,D_{c}^{3}+9\,D_{d}^{3}\,.
\end{equation}
The K\"ahler
cone is the subspace of the space of parameters $t_{i}$ for which
the condition $\int_{C}J>0$ holds. In this case, the
K\"ahler cone conditions are:
\begin{eqnarray}
t_1-2\,t_3 & > & 0\,,\nonumber \\
-2\,t_1+t_2+3\,t_3 & > & 0\,,\label{kkfirstmodel}\\
t_2-3\,t_4 & > & 0\,,\nonumber \\
2\,(t_3-t_2)+t_4 & > & 0\,.\nonumber 
\end{eqnarray}

Now that we have the volume \eqref{volfirstmodel} in explicit `Swiss cheese' form, we can search for the large volume limit at which we would like to stabilize the CY. The idea is to find the right divisor $\eta_l$, such that when its volume $\tau_l$ grows, only $\tau_a$ will grow, and $\tau_b, \ldots, \tau_d$ will remain constant. In this case, $\eta_l$ is clearly $\eta_3 = D_8$. Na\"ively, we could declare our large volume limit to be
\begin{equation}
{\rm Naively:} \quad \tau_3 \rightarrow \infty\,; \quad \tau_1\,, \tau_2\,, \tau_4 \quad {\rm constant \ and \ small}\,.
\end{equation}

By looking at the projective weights of the coordinates in table \ref{chargesfirstmodel}, we conclude that any divisor that is charged w.r.t. the third row will grow large, whereas any divisor that is not will remain constant in volume. Henceforth, we will refer to $\eta_3$ as a `large direction', or `large' divisor. However, care must be exercised in trying to shrink the so-called `small' divisors. Although one would, by inspection of \eqref{volfirstmodel}, conclude, that the directions $\tau_1, \tau_2,$ and $\tau_4$ can be shrunk to arbitrarily small size while keeping $\tau_3$ arbitrarily large, a careful analysis of the K\"ahler cone conditions \eqref{kkfirstmodel} reveals that this is not entirely possible. If we rewrite these conditions in terms of the divisor volumes as follows:
\begin{eqnarray*}
7\sqrt{\tau_{b}}-3\sqrt{\tau_{c}} & > & 0\,,\\
3\sqrt{\tau_{d}} & > & 0\,,\\
\sqrt{\tau_{a}}-5\sqrt{\tau_{b}} & > & 0\,,\\
-\sqrt{\tau_{a}}+5\sqrt{\tau_{b}}+5\sqrt{\tau_{c}}-\sqrt{\tau_{d}} & > & 0\, ,
\end{eqnarray*}
we see from the last condition that sending $\tau_a$ large forbids setting both $\tau_b$ and $\tau_c$ very small. At least one of these two volumes will have to be large. By carefully analyzing these conditions, we conclude that the only possible large volume limits are the following two:
\begin{eqnarray}
&\tau_1&\,, \tau_4\, \rightarrow 0\,, \quad \tau_2\,, \tau_3 \, \rightarrow \infty\,,\\
{\rm and} \quad &\tau_2&\,, \tau_4\, \rightarrow 0\,, \quad \tau_1\,, \tau_3 \, \rightarrow \infty\,.
\label{lvlimfirst}
\end{eqnarray}
As we will see in the next subsection, this phenomenon can be linked to the topology of the divisors.

\subsubsection{Identifying smooth, `small' cycles}
We will now search for all smooth, potentially `small', effective divisors in this model, on which we will subsequently wrap our MSSM branes and our E3-branes. We will require smoothness, in order to be able to reliably compute Hodge numbers and induced charges.

As explained in the previous section, any divisor that is not charged under the third $\mathbb{C}^*$ action shown in table \ref{chargesfirstmodel} has at least the potential to be `small'. In other words, such a divisor must be of the form $D = k\,\eta_1+l\,\eta_2+m\,\eta_4$.  However, by inspecting the weight table, we see that such a divisor will always only have one monomial to represent it, namely
\begin{equation}
x_5^k\, x_6^l\, x_8^m\,.
\end{equation}
Hence, the only smooth (i.e. irreducible), small divisors are $D_5$, $D_6$, and $D_8$.\\

We would now like to compute the Hodge numbers of these three divisors. Given the fact that all three of them are rigid (i.e. have no deformations), Serre duality tells us that they have $h^{0,2} = 0$. In order to compute $h^{0,1}$, we will use the index formula for the holomorphic Euler characteristic \eqref{holomorphiceuler}. Plugging in the data for this CY, we find for $D = k\,\eta_1+l\,\eta_2+m\,\eta_4$
\begin{eqnarray*}
\chi\left(D,\mathcal{O}_{D}\right) & = & -\frac{1}{3}\,k + \frac{4}{3}\, k^3 - \frac{1}{3}\,l + \frac{3}{2}\, k^2 l - \frac{5}{2}\, k l^2 + \frac{4}{3}\, l^3\\ 
& & + 4\,m - \frac{21}{2}\, k^2 m + \frac{45}{2}\, k m^2 - 16\, m^3 - \frac{1}{2}\,n + \frac{1}{2}\,l^2 n - \frac{3}{2} l n^2 + \frac{3}{2}\, n^3\,.
\end{eqnarray*}

Looking for a choice of parameters $\left(k, l,m \right)$ such that $\chi\left(D,\mathcal{O}_{D}\right)=1$ 
we find the solutions\[
\left(k,l,m\right)=\left\{ \left(1,0,0\right),\left(0,1,0\right),\left(0,0,1\right),\left(1,1,0\right),\left(0,1,1\right),\left(1,1,1\right)\right\} .\]
The last three divisors in the list are reducible, and hence not smooth. The first three are precisely the ones we identified before. This calculation shows that all three of them have $h^{0,1}=0$, i.e. no Wilson lines. This means that these divisors are perfect for all our purposes: We want to avoid having extra neutral zero-modes on the instantons, we do not want to have D-branes with extra moduli to stabilize, and we want to be able to turn on NSNS three-form flux without causing any Freed-Witten anomalies, all of which is avoided by having $h^{0,2}=b^1=b^3=0$.\\

We can actually identify these divisors as \emph{rational surfaces}. Rational surfaces are either Hirzebruch surfaces, $\mathbb{CP}^2$, or blow-ups of $\mathbb{CP}^2$ at up to eight points. First of all, we notice that for all three $D_5, D_6, D_8$, the second plurigenus vanishes
\begin{equation}
p^2(D) \equiv {\rm dim}\, H^0(D, K_D^{\otimes 2})={\rm dim}\, H^0(D, N_D^{\otimes 2}) = 0\,,
\end{equation}
where $N_D$ is the normal bundle of $D$. We see this by inspecting the table \ref{chargesfirstmodel}, and seeing that, for instance, a section of ${N_{D_5}}^{\otimes 2}$ would correspond to a monomial of class $2\,\eta_1$ that does not vanish on $D_5$. The only monomial in this class is $x_5^2$,\footnote{Checking this is not entirely trivial. One must also take the SR ideal in \eqref{srfirstmodel} into account. As certain monomials are not allowed to vanish on a surface, it is possible to build sections that are quotients of monomials.} so there are no non-vanishing sections of this bundle. The same occurs for the other two divisors. 
The vanishing of the second plurigenus, plus the fact that $h^{0,1}=0$, implies by the Castelnuovo-Enriques Theorem (see section 4.4 of \cite{GH}) that these surfaces are rational.
The Euler numbers of the three divisors are easily computed by means of the formula $\chi(D) = D^3+c_2(X) \cdot D$ to be
\begin{eqnarray*}
\chi(\eta_1) &=& 4\,, \quad \chi(\eta_2) = 4\,, \quad \chi(\eta_4) = 3\,,\\
\Rightarrow \quad h^{1,1}(\eta_1) &=& 2\,, \quad h^{1,1}(\eta_2) = 2\,, \quad h^{1,1}(\eta_4) = 1\,.
\end{eqnarray*}
Let us take a closer look at $D_5$. By inspecting \eqref{srfirstmodel}, we see that both $x_1$ and $x_8$ must be non-vanishing. Hence, we can gauge-fix both coordinates $x_1=x_8=1$. This uses up two projective $\mathbb{C}^*$ actions. Let us choose the gauge fixing such that the first and the last rows of table \ref{chargesfirstmodel} are eliminated. If we now write down the polynomial defining the CY, after setting $x_5=0$ and gauge-fixing, we have something of the form: $P(x_2, x_3, x_4, x_6, x_8) + x_7=0$. Hence, $x_7$ is uniquely determined by the other coordinates, so we can eliminate it. After taking the appropriate linear combination of the charge rows, we are left with the following toric description of the surface given in table \ref{tab:hirzebruch}:
\begin{table}[h]
\begin{centering}
\begin{tabular}{|c|c|c|c|}
\hline 
$x_{2}$ & $x_{3}$ & $x_{4}$ & $x_{6}$\tabularnewline
\hline 
\hline
6 & 1 & 1 & 1\tabularnewline
\hline 
1 & 0 & 0 & 1\tabularnewline
\hline
\end{tabular}
\par\end{centering}

\caption{charges}  \label{tab:hirzebruch}
\end{table}

\noindent This is nothing other than the fifth Hirzebruch surface $\mathbb{F}_5$. It is not a Del Pezzo surface, because its anticanonical bundle is not ample. In fact, we could have seen this more quickly by inspecting the intersection numbers \eqref{intfirstmodel}. It is well known that an ample line bundle on a surface has to have a positive intersection with any effective curve on the surface. In our case, this means that
\begin{equation}
\int_{D_5} (-K_{D_5}) \cdot C >0 \quad {\rm for \ any \ effective \ curve } \quad C \in D_5\,.
\end{equation}
Taking the curve defined as $C: \{x_5=0\} \cap \{x_6=0\}$, we can compute
\begin{eqnarray}
\int_{D_5} (-K_{D_5}) \cdot C &=& \int_{D_5} (-\eta_1) \cdot C\\
&=& \eta_1 \cdot (-\eta_1) \cdot \eta_2 = -3\,.
\end{eqnarray}
Hence, we see that this surface cannot be Del Pezzo. Similarly, we see from the number $\eta_2^2\,\eta_4=+1$ that our second surface, $D_6$ must also be a non-Del Pezzo Hirzebruch surface. This explains why we cannot simultaneously shrink both of these surfaces arbitrarily as we na\"ively would have expected. Our third surface, $D_8$, however, is simply a $\mathbb{CP}^2$, which is a Del Pezzo. It can be shrunk arbitrarily. Notice that $D_8$ is the only surface whose volume appears `diagonally' in the volume function of the CY.

\subsection{Scenarios in the first model}
\subsubsection{Step one: `Local' models}

We will study a setup with two stacks of D7 branes ${\rm D7}_A$ and ${\rm D7}_B$, each one on a different `small' cycle, plus one E3-brane on another `small' cycle. The reason for placing the MSSM on `small' cycles, is to keep the gauge coupling constants large. We want two different D7 stacks in order to get chiral matter in four dimensions. We would like the MSSM gauge group to be unitary. There are two ways to accomplish this. One way would be to have D7/image D7-brane configurations (as opposed to D7-branes on top of the O7-plane, or transversally involution invariant D7-branes). However, since the cycles we are dealing with are rigid, they are automatically left invariant by involutions of the type we consider in this paper. The other way to get unitary gauge groups is to have transversally invariant, even ranked stacks, which will induce symplectic gauge groups, and then turn on a diagonal flux to break the latter to unitary groups. 

The E3-brane on the other hand, must have an $O(1)$ gauge group. This is accomplished by having a single E3 placed on a transversally invariant cycle. 

We have three possible cycles on which to place the E3 instanton. Having fixed that choice, the two MSSM branes will occupy the other two `small' cycles. Let us begin by putting an E3 on $\eta_1$. The charge vector for this brane is
\begin{equation}
\G_{E3} = \eta_1 + \tfrac{1}{2}\,\eta_1^2 + \tfrac{7}{6}\,\omega\,,
\end{equation}
where the two-form, four-form and volume-form correspond to D3, D1, and D(-1) charges, respectively. The half-integral four-form corresponds to the flux $F = \tfrac{1}{2}\,\eta_1$ that compensates for the Freed-Witten anomaly. The four-form can be geometrically interpreted as the Poincar\'e dual to the curve on which the induced D1 would be wrapped. However, if we integrate all possible basis elements of $H^2(X, \Z)$
\begin{eqnarray}
\int_{E3} \tfrac{1}{2}\,\imath^*(\eta_1)\, \cdot \imath^*\{\eta_1, \eta_2, \eta_3, \eta_4\} &=& \int_X \tfrac{1}{2}\,\eta_1^2\, \cdot \{\eta_1, \eta_2, \eta_3, \eta_4\}\\ 
&=& \{4, \tfrac{3}{2}, -\tfrac{21}{2}, 0\}\,.
\end{eqnarray}
we see that this curve is not a well-defined (integral) element of $H_2(X, \Z)$. In other words, it fails the test we defined in section \ref{sec:deffreedwitten}, which means that this half-integer flux cannot be compensated by turning on more flux on the E3.

Due to the non-vanishing $U(1)$ worldvolume field-strength $F = \tfrac{1}{2}\,\eta_1$, this E3-brane is not invariant under orientifolding. In order to fix this we must turn on an appropriate $B$-field\footnote{Note that, because we only consider involutions with $H^2_-(X, \Z) = 0$, the $B$-field is frozen. However, it is allowed to take on discrete values such that $B = -\sigma^*(B)$ mod $H^2(X, \Z)$.}

\begin{eqnarray}
B &=& F = \tfrac{1}{2}\,\eta_1\,, \quad {\rm such \ that}\\ 
\mathcal{F} &=& F-B = 0 = -\sigma^*(\mathcal{F})\,.
\end{eqnarray}
Two comments are in order: First of all, notice, that since we now have $B = F$ on the E3-brane, the latter automatically satisfies the D-term constraint, i.e. has a vanishing FI parameter \eqref{fidefinition}. Its central charge is aligned with that of the O7-plane. Since the $B$-field cannot run continuously, this means that this instanton cannot become non-BPS (unless, of course, supersymmetry is broken by the other branes present), and we do not have to worry about extra fermionic zero-modes appearing in different regions of moduli space. As explained in \cite{GarciaEtxebarria:2007zv, GarciaEtxebarria:2008pi}, this means that this instanton will contribute to the superpotential, as opposed to giving rise to the higher F-terms considered in \cite{Beasley:2004ys, Beasley:2005iu}.
Secondly, having fixed the $B$-field at this value, it is now impossible for other instantons wrapped on the other two small cycles to contribute, as their respective Freed-Witten compensating fluxes differ from the one in this case.\\

Now, we move on to set up our MSSM D7-branes. We will do this in two stages. First, we will place two rank one D7-branes on the two remaining small divisors and tackle the problem of the unwanted E3-D7 strings.\footnote{As this problem is insensitive to the ranks of the stacks, we will set them to one for now, and adjust them later as needed. Everything we do now will carry over to the case of higher rank stacks. One just needs to take the tensor product of the line bundles we construct here with traceless vector bundles.} In the next subsection, we will scan for involutions and try to embed the system into a consistent, global (tadpole canceling) model, and see whether we can still solve the problem of unwanted charged zero-modes and unwanted matter after we are forced to add tadpole canceling `hidden' D7-branes.  

We place two D7-branes, ${\rm D7}_A$ and ${\rm D7}_B$ on the remaining small divisors, $\eta_4$ and $\eta_2$, respectively.  Both branes fail our test for the Freed-Witten flux, i.e. their FW fluxes cannot be turned off. By inspecting \eqref{intfirstmodel}, we see that $\eta_1$ and $\eta_4$ never intersect on the CY. Hence, there are no zero-modes charged under the D$7_A$. If we now compute the chiral intersection number between the E3 and the D$7_B$, as defined in equation \eqref{dsz}, we find
\begin{equation}
\langle \Gamma_{E3}\, , \, \Gamma_{B} \rangle = 4\,.
\end{equation}
This will induce four unwanted charged zero-modes in four dimensions. Hence, we need to turn on extra flux on both branes to cancel this. 

Let us define the added fluxes (on top of the half-integral fluxes) $\Delta F_B$ and $\Delta F_{E3}$ on the ${\rm D7}_B$ and the E3 as follows:
\begin{eqnarray}
\Delta F_B &=& \{b_1\,; b_2\,;b_3\,; b_4\}\,, \\
\Delta F_{E3} &=& \{e_1\,; e_2\,; e_3\,; e_4 \} \,,
\end{eqnarray}
where the $b_i$'s and $e_i$'s are integer components w.r.t the $\eta_i$ basis, and we have suppressed the pull-back symbol. Computing the charge vectors again we get the intersection number
\begin{equation}
\langle \Gamma_B\,, \, \Gamma_{E3} \rangle  = 4 - 3\,b_1 + 5\,b_2 + 3\,e_1 - 5\,e_2\,.
\end{equation}
Setting this to zero yields the following seven-parameter solution:
\begin{eqnarray}
\Delta F_B &=& \{3+e_1+5\,n\,; 1+e_2+3\,n\,; b_3\,; b_4\}\,,\\
\Delta F_{E3} &=& \{e_1\,; e_2\,; e_3\,; e_4 \} \,,
\end{eqnarray}
where $n$ is an arbitrary integer, as are the other parameters. In order to maintain the orientifold invariance of the E3-brane, the $B$-field must always be adjusted such that $B = \tfrac{1}{2}\,\eta_1+\Delta\,F_{E3}$. The number of chiral bifundamental A\,B strings is then given by
\begin{equation}
\langle \Gamma_A\,, \, \Gamma_{B} \rangle = 3\,(n+a_4-b_4)-a_2+e_2\,.
\end{equation}

For both the D$7_A$ and the D$7_B$ we get
\begin{equation}
 \xi_A, \xi_B \sim (t_2-3\,t_4) = \sqrt{\tau_4}\,. 
\end{equation} 
Hence, these D-terms want to shrink the D$7_A$ to zero size, bringing us to the boundary of the K\"ahler cone. The formula used to compute these D-terms, however, is only valid at large radius. Once the cycle $\eta_4$ reaches stringy scale, worldsheet instanton corrections will dominate and drastically modify the central charge of the D-brane. 
Computing these corrections exactly would require solving the Picard-Fuchs equations for the mirror CY, which is beyond the scope of this paper. On the other hand, it is more plausible that the cycles of both the D$7_A$ and D$7_B$ will get stabilized within the K\"ahler cone by string loop effects, as has been worked out in general in \cite{Cicoli:2007xp, Cicoli:2008va}.

Now, let us reshuffle the branes and place the E3, D7$_A$ and D7$_B$ on $\eta_4, \eta_1$ and $\eta_2$, respectively. We obtain the following solution:
\begin{eqnarray}
\Delta F_A &&\quad {\rm arbitrary}\,,\\
\Delta F_B &=& \{b_1;1+e_2+3\,n\,; b_3\,; 1+e_4+n\}\,,\\
\Delta F_{E3} &=& \{e_1\,; e_2\,; e_3\,; e_4 \} \,.
\end{eqnarray}
This system has 
\begin{equation}
\langle \Gamma_A\,, \, \Gamma_{B} \rangle = 3\,(-a_1+b_1-5\,n-3)+5\,(a_2-e_2)\,,
\end{equation}
bifundamental, chiral, A\,B strings. 

\begin{table}[h]
\begin{centering}
\begin{tabular}{|c|c|c|c|}
\hline 
Scenario&E3 & D7$_A$ & D7$_B$
\tabularnewline
\hline
\hline 
I&$\eta_1$ & $\eta_4$ & $\eta_2$ \tabularnewline
\hline 
&arbitrary & arbitrary & $\{3+e_1+5\,n\,; 1+e_2+3\,n\,; b_3\,; b_4\}$   \tabularnewline
\hline 
\hline
\hline
II&$\eta_4$ & $\eta_1$ & $\eta_2$ \tabularnewline
\hline 
&arbitrary & arbitrary & $\{b_1;1+e_2+3\,n\,; b_3\,; 1+e_4+n\}$ \tabularnewline
\hline
\end{tabular}
\par\end{centering}
\caption{Two `local' models.}
\label{tab:scenariosfirstmodel}
\end{table}

Finally, we could now go on to reshuffle the branes again, but this would force us to put the MSSM branes on $\eta_1$ and $\eta_4$, which do not intersect at all. This would defeat the purpose of having a chiral MSSM setup. We summarize the results in table \ref{tab:scenariosfirstmodel}.

\subsubsection{Step two: `Global' models}
By `global' model, we will mean a model where an involution has been chosen, and all D7-charge has been cancelled. In the previous section we identified the divisors on which we want to wrap the instanton and two stacks of intersecting MSSM branes. However, such setups will typically not cancel the total D7 tadpole, and a third (set of) branes will have to be added. It is phenomenologically desirable that these new branes do not intersect the MSSM branes, nor the E3-brane. In this section, we will find out to what extent it is possible to solve this problem.

Let us begin with scenario $I$ in table \ref{tab:scenariosfirstmodel}, where the two stacks have ranks $N_A$ and $N_B$. We will pick an involution and explain the procedure by working out the example. Define the involution as $x_1\rightarrow -x_1$. The resulting O$7$-plane has D$7$-charge 
\begin{equation}
-8 \times [{\rm O}7] = -(120\, \eta_1+72\,\eta_2+56\,\eta_3+24\,\eta_4)\,.
\end{equation}
Taking into account the D$7_A$ and D$7_B$ with their arbitrary ranks, $N_A, N_B$ and their image branes, means that we have to make up for 
\begin{equation}
120\, \eta_1+(72-N_B)\,\eta_2+56\,\eta_3+(24-2\,N_A)\,\eta_4\, \label{compensatefmfsfi}
\end{equation}
worth of D$7$-charge.
This charge can be distributed in many ways: We can distribute it among several branes or use just one brane; we can use brane/image brane pairs, or Whitney-type branes (see section \ref{sec:orientifolding}). Let us first address the question as to whether one should distribute the charge among several branes, or just a single brane. Picking a single brane with the full charge in \eqref{compensatefmfsfi} has several advantages over partitioning the charge among more branes. First of all, the hidden brane has to have zero intersection product with the E$3$, the D$7_A$ and the D$7_B$. If we were to partition the hidden brane into several branes, 
\begin{equation}
\Gamma_H = \sum_i \Gamma_{H_i}
\end{equation}
with each $\Gamma_{H_i}$ satisfying the zero intersection property, then the sum $\Gamma_H$ would also satisfy it. It is a necessary condition that the total charge satisfy it, in order to solve the problem for the constituents. It is therefore much simpler to only have to solve this problem once for one brane. The second advantage lies in the fact that a single high charge D$7$-brane will typically generate a much larger curvature induced D$3$-charge than several low charge branes. Schematically, in a one-modulus CY, the Euler number of a degree $N$ divisor grows like $\sim N^3$, whereas $N$ degree one branes will simply induce a total charge $N$. We will therefore work with a single hidden brane.

The second question concerns the type of brane we should use. We claim that it is more advantageous to use a Whitney-type hidden brane. Whitney-type branes are invariant under the involution by construction. This means that the E$3$ is automatically orthogonal to it. This also means that imposing that the hidden brane have to be orthogonal to the D$7_A$ and D$7_B$ automatically makes it orthogonal to their respective images. Finally, the invariance means that the hidden brane automatically has a trivial D-term. The D-term for a non-invariant hidden brane, which is always wrapped on a large cycle, would typically ruin the large volume limit. This can be understood as follows. The charge vector of the hidden brane is orthogonal to those of the E$3$, the two MSSM branes, and their respective images. Combining these equations, and using the diagonal basis from \eqref{diagfirstmodel} one can show that
\begin{equation}
D_{H} \cdot D_{E3, \,{\rm D}7_A, \,{\rm D}7_B} \cdot (F_H-B)=0 \Rightarrow D_{H} \cdot D_{b, \,c, \,d} \cdot (F_H-B)=0\,.
\end{equation}
Hence, the FI term for the hidden brane will necessarily be proportional to $\tau_a \sim t_3$, i.e. the volume of the large cycle. This would force the CY to be small. A Whitney-type brane automatically circumvents this problem.\\

Therefore, we will search for a single Whitney-type D$7$-brane of charge $[D_W]=2\,[D_P]$ given by \eqref{compensatefmfsfi}. The easiest way to construct its charge vector is by using the K-theoretic picture, as described in \cite{Collinucci:2008pf} summarized in appendix \ref{sec:k-theory}. As explained there, the choice of shift flux does not enter the intersection numbers between the Whitney brane and the other branes present. All to do is to solve the equations:
\begin{equation}
\langle \Gamma_H\,, \, \Gamma_A \rangle = \langle \Gamma_H\,, \, \Gamma_B \rangle=0\,.
\end{equation}
The solution is simply $N_B = 3\,N_A$. This means that this scenario can generate models with gauge groups of the form $U(3\,N) \times U(N)$. Let us now compute the tadpole that our hidden brane generates. For this computation, we will have to assign a value to the shift vector. For simplicity, let us choose $N_B = 3, N_A=1$. The constraints on the shift flux $S$ from \eqref{whitneyconstraints} become
\begin{equation*}
\{52-e_1, \tfrac{57}{2}-e_2, \tfrac{49}{2}-e_3, \tfrac{19}{2}-e_4\} \geq S \geq \{7-e_1, \tfrac{9}{2}-e_2, \tfrac{7}{2}-e_3, \tfrac{3}{2}-e_4\}\,.
\end{equation*}
Notice that we cannot saturate these constraints. This might indicate the presence of a flux on the brane that cannot be switched off due to some anomaly. Let us choose $S$ to be `minimal',
\begin{equation}
S =  \{7-e_1, 5-e_2, 4-e_3, 2-e_4\}\,.
\end{equation}
Now we may compute the `physical' (gauge invariant) D3 tadpole by taking the six-form component as follows:
\begin{equation}
\Big(\Gamma_W\,e^{-B} \Big)_{6-{\rm form}} = \frac{7763}{4} \approx 1940\,.
\end{equation}

Let us now repeat this calculation for the second scenario. Starting with arbitrary ranks $N_A$ and $N_B$ again, we solve the equations
\begin{equation}
\langle \Gamma_W\, , \Gamma_A \rangle=0\,, \quad \langle \Gamma_W\, , \Gamma_B \rangle=0\,.
\end{equation}
The second equation is proportional to $3\,N_A-5\,N_B$. Setting it to zero and eliminating $N_A$ in the first equation yields a term proportional to $N_B$. This means, we cannot choose non-zero ranks and turn off the intersections with the hidden brane. Therefore, the second scenario has a visible `hidden' sector.

\subsection{Moduli stabilization analysis}

From equation (\ref{equ:d-term-general}) we see that as long as the magnetized D7-branes are small the potential is of order $1/{\cal V}^2$, so the D-term part will dominate over the F-term contribution in the LVS. The curvature along its non-flat directions is much larger than the one of the F-term potential. In the limit of exponentially large volume (the divisor of the D7-brane has to remain small) this generates an exponentially strong force in comparison to the F-term forces. Hence, in the following, we will use $V_D=0$ as a constraint on our configuration and just look at the F-term potential.

To obtain the concrete form of the F-term potential (\ref{equ:f-term}) for our scenarios we have to calculate the self-intersection volume for the instanton. In the first case it is given by
\begin{equation}
 {\rm Vol}\left(D_{E3}\cap D_{E3}\right) = 8 t_1 +3 t_2-21t_3=-\frac{\sqrt{2}}{5}\left(7\sqrt{\tau_b}+3\sqrt{\tau_c}\right)\,,
\end{equation}
and in the second one we obtain
\begin{equation}
 {\rm Vol}\left(D_{E3}\cap D_{E3}\right) = -3 t_2+9 t_4=-3\sqrt{2}\sqrt{\tau_d}\,.
\end{equation}
Knowing these, we can write the potentials as a function depending on $\tau_a,\,\tau_c,\,\tau_d$ and $\tau_{E3}$. So for the first scenario we find
\begin{eqnarray}
 V_F&=& \frac{1}{\hat{\cal V}^2}\Big(  \frac{\sqrt{2}}{5} 4 \pi^2 \left(7\sqrt{5\tau_{E3}+\tau_c}+3\sqrt{\tau_c}\right)  \hat{\cal V}\left|A_{E3}\right|^2 e^{-4\pi\hat\tau_{E3}} \nonumber\\
& &-4\pi \hat\tau_{E3} e^{-2\pi \hat\tau_{E3}} \left|A_{E3} W_0\right|+\frac{3}{4}\frac{\hat\xi}{\hat{\cal V}}\left| W_0\right|^2\Big)\, ,
\end{eqnarray}
where ${\cal V}$ is also a function of the divisor volumes above.
Now we search for a minimum of the potential 
\begin{equation}
 {\rm d} V_F=0 \quad \Rightarrow \quad \frac{\partial V_F}{\partial \tau_d}=\frac{\partial V_F}{\partial \hat {\cal V}}\frac{\partial \hat {\cal V}}{\partial \tau_d}=0\,.
\end{equation}
Hence, we can also solve $\frac{\partial V_F}{\partial \hat {\cal V}}=0$. 
\begin{eqnarray}
\Rightarrow \quad {\cal V}&=&\frac{5 g_s  W_0 \tau_{E3} e^{\frac{2 \pi \tau_{E3}}{g_s}}  }{
 \sqrt{2}A_{E3} \pi (3\tau_c + 7  \sqrt{5 \tau_{E3} + \tau_c})} \nonumber \\ 
& &\pm \frac{\sqrt{5}g_s  W_0 e^{\frac{2 \pi \tau_{E3}}{g_s}}\sqrt{80   \tau_{E3}^2 - 9\sqrt{2} \xi (3  \sqrt{\tau_c} + 7  \sqrt{5 \tau_{E3} + \tau_c}})}{
 4\sqrt{2}A_{E3} \pi (3\tau_c + 7  \sqrt{5 \tau_{E3} + \tau_c})}\,.
\label{equ:prob-f-term}
\end{eqnarray}
What is important here is that, although the potential looks like the one in \cite{Blumenhagen:2007sm}, there is a subtle difference to our case. If we demand a large volume while at the same time fulfilling the K\"ahler cone constraints, the term in the square root becomes negative. Hence we cannot realize the desired LVS in scenario I.\footnote{This confirmes the theorem  of \cite{Cicoli:2008va} that one obtains a minimum at exponentially large volume only if the instanton is wrapped around a local blow-up mode resolving a point-like singularity.}

Let us now look at the second scenario. Here we obtain the same form for the potential as~\cite{Balasubramanian:2005zx}
\begin{eqnarray}
 V_F&=& \frac{1}{\hat{\cal V}^2}\Big(  12 \pi^2 \sqrt{2\tau_{E3}}  \hat{\cal V}\left|A_{E3}\right|^2 e^{-4\pi\hat\tau_{E3}}\nonumber\\
& &-4\pi \hat\tau_{E3} e^{-2\pi \hat\tau_{E3}} \left|A_{E3} W_0\right|+\frac{3}{4}\frac{\hat\xi}{\hat{\cal V}}\left| W_0\right|^2\Big)\, ,\label{equ:d-pot-dia}
\end{eqnarray}
so the only thing that could prevent us from having large volume stabilization are the the K\"ahler cone (KC) conditions. This means that in the second scenario, although we could not solve the intersection problem, we can at least realize the LVS. Looking at the allowed large volume limits in \eqref{lvlimfirst}, we can pick the case where $\tau_2$ and $\tau_4$ small. In this case, we have to drop the D7$_A$-brane, since it would generate an FI-term of the form $\sqrt{5\,\tau_1+3\,\tau_2+\tau_4}$, which would be fatal to this LVS. Choosing~$|A_{E3}|=1$, $|W_0|=5$ and $g_s=\frac{1}{10}$ we find the following values for the CY and instanton volume at the location of the minimum
\begin{eqnarray*}
\tau_{E3} & = & 2.15\,,\\
{\cal V} & = & 1.46514 \cdot 10^{57}\, .
\end{eqnarray*}
Note, however, that the FI-term generated by the remaining D7$_B$-brane, which is of the form $\sqrt{3\,\tau_2+\tau_4}$, would na\"ively force the instanton cycle to zero size, thereby destroying this LVS. It is, however, possible that the string loop corrections considered in \cite{Cicoli:2007xp} might counter this effect and keep the instanton size finite. But this is beyond the scope of our paper.

The fact that the volume of $\eta_4$ appears diagonally in the CY volume is enough to get the right form of the F-term potential. One can also show that this surface actually resolves a point-like singularity. This is another affirmation of the theorem given in~\cite{Cicoli:2008va}. 
\section{Second model} \label{sec:T4model}

The results for all of our scenarios throughout the paper are concisely summarized in \mbox{table \ref{tab:finalresults}}.
\subsection{R1 resolution of $\mathbb{P}^4_{1,1,2,2,6}\left( 12 \right) /\mathbb{Z}_2:1\,0\,1\,0\,0$
geometry}
Our next model is the first of two resolutions of the orbifolded weighted projective space $\mathbb{P}^4_{1,1,2,2,6}\left( 12 \right) /\mathbb{Z}_2:1\,0\,1\,0\,0$. Here, the integers $(1, 0, 1, 0, 0)$ denote the charges of the coordinates under the $\mathbb{Z}_2$ action.\\
The projective weights for this model are given by table \ref{tab:t4charges}.

\begin{table}[h!]
\begin{centering}
\begin{tabular}{|c|c|c|c|c|c|c|c||c|}
\hline 
$x_{1}$ & $x_{2}$ & $x_{3}$ & $x_{4}$ & $x_{5}$ & $x_{6}$ & $x_{7}$ & $x_{8}$ & \emph{p}\tabularnewline
\hline
\hline 
2 & 1 & 6 & 1 & 2 & 0 & 0 & 0 & 12\tabularnewline
\hline 
2 & 1 & 5 & 0 & 2 & 0 & 0 & 2 & 12\tabularnewline
\hline 
2 & 0 & 5 & 1 & 2 & 0 & 2 & 0 & 12\tabularnewline
\hline 
1 & 0 & 3 & 0 & 1 & 1 & 0 & 0 & 6\tabularnewline
\hline
\end{tabular}
\par
\end{centering}
\caption{Projective weights for the R1 resolution of $\mathbb{P}^4_{2,1,6,1,2}\left( 12 \right) /\mathbb{Z}_2:0\,0\,1\,1\,0$.} \label{tab:t4charges}
\end{table}

The Stanley-Reisner ideal of the ambient space reads
\begin{equation} \label{srr1}
SR=\{x_2 x_3,\,x_2 x_4,\,x_3 x_4,\,x_3 x_6,\,x_4 x_7,\,x_2 x_8,\,x_1 x_5 x_6 x_7,\, x_1 x_5 x_6 x_8,\, x_1 x_5 x_7 x_8\}\,.
\end{equation}

The triple intersection numbers in the basis $\eta_{1}=D_{2}$, $\eta_{2}=D_{4}$,
$\eta_{3}=D_{6}$, $\eta_{4}=D_{8}$ are encoded in

\begin{eqnarray*}
I_{3} & = &-78\eta_4^3 -6\eta_3 \eta_4^2 -6\eta_3^2 \eta_4 +2\eta_3^3 +36\eta_2 \eta_4^2 +6\eta_2 \eta_3 \eta_4 +\eta_2 \eta_3^2 \\& &-18\eta_2^2 \eta_4 -3\eta_2^2 \eta_3 + 9\eta_2^3+ \eta_1 \eta_3^2  -3\eta_1^2 \eta_3 +9\eta_1^3\,. 
\end{eqnarray*}

The K\"ahler form in the basis $\left\{ \eta_{1},\eta_{2},\eta_{3},\eta_{4}\right\} $
is given by\[
J=t_{1}\,\eta_{1}+t_{2}\eta_{2}+t_{3}\,\eta_{3}+t_{4}\,\eta_{4},\]
the volumes of the corresponding divisors are

\begin{eqnarray*}
\tau_{1} & = & \frac{1}{2}\left(-3 t_1 + t_3\right)^2\,,\\
\tau_{2} & = & \frac{1}{2}\left(-3 t_2 + t_3 + 6 t_4\right)^2\,,\\
\tau_{3} & = & \frac{1}{2}\left(-3 t_1^2 - 3 t_2^2 + 2 t_1 t_3 + 2 t_2 t_3 + 2 t_3^2 + 
 12 t_2 t_4 - 12 t_3 t_4 - 6 t_4^2\right)\,,\\
\tau_{4} & = & \frac{1}{2}\left(-18 t_2^2 + 12 t_2 t_3 - 6 t_3^2 + 72 t_2 t_4 - 12 t_3 t_4 - 
 78 t_4^2\right)\,.
\end{eqnarray*}

The volume of the CY manifold is given by
\begin{eqnarray} 
\mathcal{V} & = & \frac{1}{18}\left[ 9 t_4^3- (-2 t_3 + 3 t_4)^3  - (-3 t_1 + t_3)^3 - (-3 t_2 + t_3 + 6 t_4)^3  \right] \label{volft4model}\\
 & = & \frac{\sqrt{2}}{9}\left[\frac{3}{2\sqrt{6}}\left(\tau_1+5\tau_2+3\tau_3+2\tau_4\right)^\frac{3}{2}-\frac{1}{2\sqrt{2}}\left(\tau_1+\tau_2+3\tau_3\right)^\frac{3}{2}-\tau_1^\frac{3}{2}-\tau_2^\frac{3}{2}\right]. \nonumber
 \end{eqnarray} 
It has the expected Swiss cheese form. From this volume formula we
deduce the diagonal basis to be
\begin{eqnarray}\label{diagt4model}
D_{a} & = & \eta_1+5\eta_2+3\eta_3+2\eta_4\,, \nonumber\\
D_{b} & = & \eta_1+\eta_2+3\eta_3\,, \\
D_{c} & = & \eta_1\,, \nonumber \\
D_{d} & = & \eta_2\,, \nonumber
\end{eqnarray} 
and the triple intersections can be rewritten as\[
I_{3}=24\,D_{a}^{3}+72\,D_{b}^{3}+9\,D_{c}^{3}+9\,D_{d}^{3}.\]

The K\"ahler cone conditions are:\begin{eqnarray*}
t_3 - t_4 & > & 0\,,\\
t_2 - t_3 - t_4 & > & 0\,,\\
t_1 - t_3 + t_4 & > & 0\,,\\
-3 t_2 + t_3 + 6 t_4 & > & 0\,,\\
-3 t_1 + t_3 & > & 0\,.
\end{eqnarray*}

For this model, the `small', rigid cycles with holomorphic Euler characteristic are
\begin{eqnarray*}
\{ D_2,\,D_4,\,D_6  \} &=& \{\eta_1,\, \eta_2,\, \eta_3 \}\,,\\
{\rm with} \quad h^{1,1} &=& \{1\,, 1\,, 8  \}\,.
\end{eqnarray*}
The first two surfaces are necessarily $\mathbb{CP}^2, \mathbb{CP}^2$. The third one, however, cannot be a Del Pezzo due to the intersection number $\eta_3^2\,\eta_2=+1$, which implies that the anti-canonical bundle is not ample. This means that this surface must be a blow-up of $\mathbb{CP}^2$ at seven points that are not in generic position. In appendix \ref{sec:T1model}, we will explicitly work out one such `pathological' surface that also fails to be a Del Pezzo.

\subsection{Scenarios in the second model}
The divisors $\eta_1$ and $\eta_2$ do not intersect, therefore, we again only have two possible scenarios, which we summarize in table \ref{tab:scenariost4model}.

\begin{table}[h]
\begin{centering}
\begin{tabular}{|c|c|c|c|}
\hline 
Scenario&E3 & D7$_A$ & D7$_B$
\tabularnewline
\hline
\hline 
I&$\eta_1$ & $\eta_2$ & $\eta_3$ \tabularnewline
\hline 
&arbitrary & arbitrary  & $\{b_1\,;b_2\,;-2 + 3\,(b_1-e_1)+ e_3\,;b_4 \}$ \tabularnewline
\hline 
\hline
\hline
II&$\eta_2$ & $\eta_1$ & $\eta_3$ \tabularnewline
\hline 
&arbitrary & arbitrary & $\{b_1\,; b_2\,; -2+e_3+3\,(-2\,b_4-e_2+2\,e_4+b_2)\,;b_4 \}$ \tabularnewline
\hline
\end{tabular}
\par\end{centering}

\caption{Two `local' models.}
\label{tab:scenariost4model}
\end{table}

Let us move on to the global analysis. We pick, for convenience, the involution \mbox{$x_3 \rightarrow -x_3$}. Solving the equations
\begin{equation}
\langle \Gamma_W\,, \, \Gamma_A \rangle = \langle \Gamma_W\,, \, \Gamma_B \rangle=0\,,
\end{equation}
we find the following solutions:
\begin{enumerate}
\item {\bf Scenario $I$}:
The constraints we get from setting the chiral intersections with the hidden brane to zero are the following:
\begin{eqnarray}
N_A &=& 3\,N_B\,,\\ 
b_4&=& -1+2\,(b_1-e_2)+e_4\,.
\end{eqnarray}
We again have a setup that requires further constraints on the `local' model. This time, however, these constraints are particularly simple to solve. To get an idea of how much D3 tadpole this Whitney-type brane can induce, let us compute it for the `minimal' choice of the shift vector $S$ in formula \eqref{whitneyconstraints}:
\begin{equation}
Q_{W, D3} = 372-\frac{3}{2}\,N_A-21\,{N_A}^3\,.
\end{equation}
This function bears a striking similarity with the results found in the previous model in appendix \ref{sec:T1model}.
Finally, let us compute the FI-terms for both MSSM branes in light of these constraints:
\begin{equation}
\xi_A,\, \xi_B \propto \sqrt{\tau_2}\,.
\end{equation}

The self-intersection volume for the instanton in this scenario is given by
\begin{equation}
 {\rm Vol}\left(D_{E3}\cap D_{E3}\right) = 9t_1-3t_3=-3\sqrt{2}\sqrt{\tau_c}=-3\sqrt{2}\sqrt{\tau_1}\,.
\end{equation}

Writing the K\"ahler cone in the diagonal basis yields
\begin{eqnarray*}
 \sqrt{\tau_a}-3\sqrt{\tau_b} & > & 0\,,\\
2\sqrt{\tau_b}-\sqrt{\tau_d} & > & 0\,,\\
2\sqrt{\tau_b}-\sqrt{\tau_c} & > & 0\,,\\
\sqrt{\tau_d} & > & 0\,,\\
\sqrt{\tau_c} & > & 0\,.
\end{eqnarray*}
Note that these conditions imply that we are free to make all three cycles $\eta_1, \eta_2, \eta_3$ small and still have a large volume limit where $\tau_4$ is kept large.

We observe that in this scenario the D-term forces us to the boundary of the K\"ahler cone. Relaxing the K\"ahler cone relations a bit and allowing non-strict inequalities , we will use $\tau_2=0$ as a constraint in the following. The F-term potential takes the form of (\ref{equ:d-pot-dia}) with the minimum
\begin{equation}
\tau_{E3}=\tau_1=1.25\,, \quad {\cal V}= 2.5945 \cdot 10^{32}\,.
\end{equation}

\item {\bf Scenario $II$}:\\
The constraints for this scenario are the following:
\begin{eqnarray}
N_A &=& 3\,N_B\,,\\ 
b_4&=& 2\,n+1+e_4\,, \quad {\rm for } \quad n \in \mathbb{Z}\,,\\
b_2&=& 5\,n+3+e_2\,.
\end{eqnarray}
Let us also compute the D3 tadpole for this hidden brane with the `minimal' choice of $S$:
\begin{equation}
Q_{W, D3} = 372+\frac{3}{2}\,N_A-75\,N_A^3\,.
\end{equation}
The function is identical to that of the first scenario. In this case, both branes give again similar FI terms
\begin{equation}
\xi_A, \, \xi_B \propto \sqrt{\tau_1}\,.
\end{equation}
In this scenario, the self-intersection volume is
\begin{equation}
 {\rm Vol}\left(D_{E3}\cap D_{E3}\right) = 9 t_2-3 t_3-18 t_4=-3\sqrt{2}\sqrt{\tau_d}=-3\sqrt{2}\sqrt{\tau_2}\,.
\end{equation}

Here, we are again forced to the boundary of the K\"ahler cone. Again, relaxing the strict inequalities, we impose $\tau_1=0$. The F-term potential takes the form of (\ref{equ:d-pot-dia}) with the minimum:
\begin{equation}
 \tau_{E3}=\tau_2=1.25\,, \quad {\cal V}= 2.5945 \cdot 10^{32}\,.
\end{equation}

\end{enumerate}

Thus in the LVS of this model we were able to stabilize in both scenarios three out of the four K\"ahler moduli and again we expect that one can stabilize the last modulus via string loop corrections \cite{Cicoli:2008va}.\\
Note that both scenarios here yield the same potentials and the same values for the volumes. This is possibly due to the fact that both divisors $\eta_1$ and $\eta_2$ have the same topology: Both are $\mathbb{CP}^2$'s. In fact, we notice in table \eqref{tab:t4charges} and in \eqref{srr1} that the CY threefold is symmetric under the simultaneous exchanges
\begin{equation}
x_2 \leftrightarrow x_4 \quad x_7 \leftrightarrow x_8\,.
\end{equation}


\section{Conclusions}

\begin{table}[h!]
\begin{centering}
\begin{tabular}{|c|c|c|c|c|c|c|c|c|}
\hline 
Scenario& $1\,(I)$ & $1\,(II)$& $2\,(I)$& $2\,(II)$& $3\,(I)$& $3\,(II)$& $4\,(I)$& $4\,(II)$
\tabularnewline
\hline
\hline 
Global&$\surd$ &$\times$&$\surd$&$\surd$&$\surd$&$\surd$&$\times$&$\times$ \tabularnewline
\hline 
St. mod. &$\times$ &$\times$ &3 &3 &$\times$ &$\times$ &3& $3$   \tabularnewline
\hline
\end{tabular}
\par\end{centering}
\caption{Summary of results. The labels represent the model numbers and scenario numbers. For each scenario we indicate with a $\surd$ or a $\times$, whether the `global problem' of supressing undesirable intersections while canceling the D7 tadpole was solved. We also indicate how many, if any, K\"ahler moduli were succesfully stabilized in each scenario.}
\label{tab:finalresults}
\end{table} 

In this paper, we have searched for realizations of the Large Volume Scenario that are compatible with the presence of MSSM D7-branes with chiral matter, in the sense explained in \cite{Blumenhagen:2007sm}. We found that it is necessary to have at least three cycles that contribute negatively to the CY volume: Two on which we placed two D7-stacks, and one for the E3-brane. 

For this purpose we searched the list of 1197 toric CY hypersurfaces with $h_{11}=4$. For simplicity we started with the 11 simplicial polytopes, which correspond to weighted projective spaces or quotients thereof.
An extension of the package PALP \cite{Kreuzer:2002uu,wwwCY}  has been used to triangulate the 8 polytopes for which all divisors on the CY are toric and to compute the Mori cone and the intersection rings. We thus found four inequivalent CYs of large-volume type.

Properly taking into account the fact that the Freed-Witten anomaly forces most of the branes (both D7 and E3) to be magnetized, we found that requiring vanishing chiral intersections between the E3 and the D7's, and between the `hidden' D7 and the rest of the setup is stringent enough to rule out some of these models entirely. Throughout this paper, we used the representation of D7-branes in terms of D9-anti-D9 condensates, which simplifies calculations of both induced charges and chiral intersections greatly. We did not specifically count vector-like pairs of chiral modes, but this can easily be done by literally counting sections of the appropriate bundles as opposed to using index theory. These issues were also carefully considered in \cite{Blumenhagen:2008at}.

For each model, we analyzed the topologies of the rigid, complex surfaces. We found that not all surfaces that are `small', in the sense that they contribute negatively to the CY volume, are also Del Pezzo. We found, as expected, that only the surfaces that \emph{are} Del Pezzo can be shrunk arbitrarily without spoiling the desired LVS. This means, that some K\"ahler moduli cannot be stabilized by instanton effects or by D7-brane induced D-term constraints. Further analysis is needed to determine, whether string loop corrections \cite{Cicoli:2007xp} can lift those flat directions. 

Our approach can be seen as complementary to one of the approaches presented in \cite{Blumenhagen:2008at}. We are searching for CY's with Del Pezzos in them by searching for polytopes with the right properties. One of the several approaches of \cite{Blumenhagen:2008at}, which is based on the techniques developed in \cite{Grimm:2008ed}, on the other hand, was to start with a simple CY, i.e. the quintic, and subject it to Del Pezzo transitions, thereby designing the desired divisor structure.

So far, our search has only yielded $\mathbb{CP}^2$ surfaces as true Del Pezzo surfaces. This paper should be considered as step one in the search for candidate `Swiss cheese' CY's. We expect that taking into account the general rules laid out in \cite{Cicoli:2008va}, combined with the techniques presented here, will  lead to viable models.
\vskip 2mm

Our results are summarized in table \ref{tab:finalresults}. For each scenario of each of the four CY's, we state whether the `global' problem of setting up E3 and D7-branes such that only wanted intersections are present, and such that the D7 tadpole is canceled, is solved. For each scenario we also give the number of K\"ahler moduli that were successfully stabilized. If no LVS was possible, we put a cross in the slot.
\vskip 5mm
In conclusion, we have shown that even with the more stringent conditions imposed by the Freed-Witten anomaly it is still possible to combine the LVS with setups of chirally intersecting D7-branes. The constraints help rule out some models, but still allow for flexibility.
We have also demonstrated that the use of `Whitney-type' branes is preferable to the use of the more familiar stacks of brane/image-brane pairs, whenever possible, because the former do not produce unwanted chiral intersections, and they induce a lot more of the desired D3-charge.

\section*{Acknowledgements}
It is a pleasure to thank R. Blumenhagen, M. Cicoli, F. Denef, and T. Grimm for very useful discussions. We are greatly indebted to H. Skarke for helping us work out the topologies of our surfaces. This work was supported in part by the Austrian Research Funds FWF under grant numbers P19051-N16, and P18679-N16.


\appendix

\section{Definitions and rules for B-branes} \label{app:branes}
\subsection{D-brane charges} \label{sec:dbranecharges&dsz}
We will be studying D-branes wrapped on even-dimensional submanifolds of a CY threefold  $X$ in IIB theory, or \emph{B-branes}. More precisely, we will deal with spacetime filling D7- and D3-branes on one hand, and Euclidean D3-branes, henceforth referred to as E3-branes, that are point-like in four dimensions.

For a D-brane wrapped on a submanifold $P$, the coupling to the total RR-potential $C = C_0+C_2+C_4+C_6+C_8$ is given by the following:
\begin{equation}
 S^{\rm Dbrane}_{P,C} = 2 \pi \int_P C \wedge e^{-B} \, {\rm Tr} \, e^F \label{defcs}
 \sqrt{\frac{\widehat{A}(TP)}{\widehat{A}(N_P)}},
\end{equation}
where $B$ is the NSNS two-form (pulled back onto $P$), $F$ is the quantized $U(1)$ field-strength of the worldvolume theory, $\widehat A$ is the `A-roof genus', and $TP$ and $N_P$ are the tangent and normal bundles of $P$, respectively. Define a polyform $\Gamma \in H^{\rm even}(X,\Z)$ such that
\begin{equation}
 S^{\rm Dbrane}_{P,C} = 2 \pi \int_P C \wedge e^{-B} \, \Gamma\,.
 \end{equation}
Then $\Gamma$ is interpreted as a source for RR-charges, or a charge vector. 

Throughout this paper, we will neglect possible torsion charges and will always deal with divisors with $b^3=0$, such that the pullback of the NSNS three-form $H$ is trivial on our D-branes. Hence, we can define D-brane charge by means of the cohomology of the internal space $X$. The most general `charge vector' $\G$ will be of the form
\begin{equation}
\G = q_{D9} + q_{D7}^A\,D_A+ q_{D5,A}\,\tilde{D}^A + q_{D3}\,\omega\,,
\end{equation}
where the ${D_A}$ and ${\tilde{D}^A}$ define base for $H^2(X, \Z)$ and $H^4(X,\Z)$, respectively, and $\omega$ is the volume-form of $X$.\\
For instance, for a single \emph{smooth} D7-brane wrapped on a four-dimensional submanifold, i.e.\~a divisor $P$, with inclusion map
\begin{equation}
\imath: P \hookrightarrow X\,,
\end{equation}
that is Poincar\'e dual to a two-form $[D_P] \in H^2(X, \Z)$, supporting a $U(1)$ field-strenght $F$, the total charge vector will be given by the following:
\begin{equation}
\G_{D7} = [D_{P}] + \Big(\int_P F\cdot \imath^*(D_A) \Big)\,\tilde{D}^A + \Big( \tfrac{1}{2}\,\int_P F^2+\frac{\chi(P)}{24} \Big)\,\omega \,. \label{defsmoothd7charges}
\end{equation} 
In this notation, the large volume formula for the FI term induced by a D7-brane is very simple:
\begin{equation}
\xi = \Im\Big(-\frac{1}{\mathcal{V}}\,\int_X e^{-(B+i\,J)} \, \Gamma_{{\rm D}7} \Big) = \frac{1}{\mathcal{V}}\,\int_{{\rm D}7}J \cdot (F-B)\,, \label{fidefinition}
\end{equation} 
where the $F$ contains the half-integral flux.\\

For two stacks of (magnetized) branes $\G_1$ and $\G_2$, the net number of chiral bifundamental strings stretching from $\G_1$ to $\G_2$ is given by the following Dirac-Schwinger-Zwanziger intersection product:
\begin{equation}
\langle \G_1, \G_2 \rangle \equiv \int_X \G_1 \wedge \G_2^*\,, \label{dsz}
\end{equation}
where $\G^*$ is defined by flipping the sign of the 2- and 6-form components
\begin{equation}
\G^* = q_{D9} - q_{D7}^A\,D_A+ q_{D5,A}\,\tilde{D}^A - q_{D3}\,\omega\,.
\end{equation}
In the case of two intersecting D7-branes with $U(1)$ fluxes $F_1$ and $F_2$, respectively, this reduces to
\begin{equation}
\int_{D7_1 \cap D7_2} (F_2 - F_1)\,.
\end{equation}

Note that, by construction, the $\Gamma$ charge vectors project out any flux on a D7-brane that is trivial in the CY even though it may be non-trivial on the divisor. Since the DSZ product depends only on the $\Gamma$'s, this means that it gives information about the net chiral spectrum, but misses possible vector-like pairs. Given two D7-branes intersecting at a Riemann surface, it is possible to count such non-chiral fields by counting sections of the corresponding bifundamental bundle over the surface. However, this is beyond the scope of this paper.

\subsection{Orientifolding} \label{sec:orientifolding}
In this paper, we will consider holomorphic involutions of the form $x_i \mapsto -x_i$, where $x_i$ is a coordinate. Involutions of this type act trivially on the (co)homology of the threefold, which implies that $H^2_-(X,\Z) = 0$, i.e. only invariant two-forms exist. We will look for involutions of O7/O3 type.

The action on the massless closed string fields is \cite{Brunner:2003zm}
\begin{eqnarray} \label{sigmaonRRNSNS}
 C_{0,4,8} &\to& \sigma^* C_{0,4,8}, \quad C_{2,6,10} \to - \sigma^*
 C_{2,6,10}, \quad g \to \sigma^* g \,,\\
  B &\to& - \sigma^* B \quad  {\rm mod}\, H^2(X, \Z)\,.
\end{eqnarray}
where the $B$-field is only well-defined up to an integral cohomology shift.
This implies that D9 and D5 charges flip sign under this action, whereas D7 and D3 charges remain intact. The action on a worldvolume gauge field living on an orientifold-invariant D3- or D7-brane stack is \cite{Gimon:1996rq}
\begin{equation} \label{sigmaonA}
 A \to - M \sigma^*A^t M^{-1} \, ,
\end{equation}
where $M$ is a symmetric or antisymmetric matrix depending on the gauge group surviving on the D-brane. We will choose the worldsheet orientifold projection such that D7-branes wrapped on an O7-plane have $O(n)$ gauge groups. For a general brane, the gauge group is decided by counting the number DN directions w.r.t. a reference D7-brane on the O7-plane. For instance, a transversally invariant D7-brane, i.e. a brane that is mapped to itself but does not lie on top of the O7-plane, will have a symplectic gauge group. On the other hand, a pair of D7-branes that are exchanged by the involution will have a unitary group, since they are not affected by the orientifolding. A transversally invariant Euclidean D3-brane, or E3-brane, will have an orthogonal group. This is summarized in the following table:
\vskip 2mm
\begin{center}
\begin{tabular}{|c|c|c|c|}
	\hline
	D7 on O7 & transversally inv. D7 & D7/image D7 pair & transversally inv. E3\\
	\hline
	$O(n)$ & $Sp(2\,n)$ & $U(n)$ & $O(n)$\\
	\hline
\end{tabular}
\end{center}
\vskip 2mm
Transversally invariant D$7$-branes necessarily satisfy the restriction that they always intersect the O$7$-plane at double points. As explained in \cite{Collinucci:2008pf},  such branes are wrapped on divisors given by equations of the form $\eta^2+\xi^2\,\chi=0$, where $\eta$ and $\chi$ are generic polynomials of appropriate degree, and $\xi=0$ is the locus of the O$7$-plane. This mimics the equation of the \emph{Whitney umbrella}. When we use such branes, we will refer to them as \emph{Whitney-type} branes for simplicity.

For open worldsheets, the $B$-field and $U(1)$ field-strength $F$ transform as follows:
\begin{equation}
B \to -\sigma^*(B) + \Lambda\,, \quad F \to -\sigma^*(F) + \imath^*(\Lambda)\,, \quad {\rm where} \quad \Lambda \in H^2(X,\Z)\,,
\end{equation}
such that the gauge invariant combination $\mathcal{F} \equiv F- \imath^*(B)$ transforms as 
\begin{equation}
\mathcal{F} \to -\sigma^*(\mathcal{F})\,.
\end{equation}
We will use the following gauge $\Lambda = 2\,B$, such that
\begin{equation}
B \rightarrow +\sigma^*(B) \quad {\rm and} \quad F \rightarrow -\sigma^*(F)+2\,\sigma^*(B)\,.
\label{involutiongauge}
\end{equation}
Throughout this paper, we will consider involutions that act trivially on the even cohomology of $X$, i.e. $h^{1,1}_-=0$. Hence, the `$\sigma^*$' can be dropped.

Given a fixed point locus of the involution consisting of O7- and O3-planes, the total charge vector will be:
\begin{eqnarray}
\G &=& \G_{O7} + \G_{O3}\,,\\
&=& [D_{O7}] + \frac{\chi(O7)}{24}\,\omega + N_{O3}\,.
\end{eqnarray}

\subsection{K-theory construction of D$7$-branes} \label{sec:k-theory}
In this appendix, we concisely explain how to describe D$7$-branes using the picture developed in \cite{Collinucci:2008pf}. There are two types of D$7$-branes we wish to describe: D$7$/image-D$7$ pairs, and transversally invariant, \emph{Whitney-type} branes. 

\subsubsection{Brane/image-brane pairs}
Let us begin by the former. Suppose we want to write the charge vector $\Gamma_{{\rm D}7}$ of a D$7$-brane wrapped on the divisor $D_P$, and its orientifold image $\Gamma_{{\rm D}7}'$, which is wrapped on a divisor in the same homology class.\footnote{Throughout this paper, we work with involutions such that $h^{1,1}_-(X)=0$.} We introduce two D$9$/$\overline{{\rm D}9}$ pairs with fluxes as follows
\begin{align}
&{\rm D9}_1: F_1=D_P-S\,, \qquad &&  \overline{{\rm D}9_1}: F_1' = S-D_P+2\,B\,,\\ 
&{\rm D9}_2: F_2=S+2\,B\,, \qquad &&  \overline{{\rm D}9_2}: F_2' = -S\,,\nonumber
\end{align}
where $S \in H^2(X, \mathbb{Z})$. The respective charge vectors $\Gamma_1, \Gamma_1', \Gamma_2, \Gamma_2'$ are expressed as follows
\begin{equation}
\Gamma = {\rm ch}(F)\,\sqrt{{\rm td}(X)} = {\rm ch}(F)\,\Big(1+\frac{c_2(X)}{24}\Big)\,.
\end{equation}
Now, we can simply write
\begin{eqnarray}
\Gamma_{{\rm D}7} &=& \Gamma_1+ \Gamma_2'\,, \nonumber \\
&=& \Big({\rm ch}(F_1)-{\rm ch}(F_2') \Big)\,\Big( 1+ \frac{c_2(X)}{24} \Big)\,.
\end{eqnarray}
It is easy to see that this expression has vanishing D$9$-charge, and that its D$7$-charge is equal to $F_1-F_2' = D_P$, as desired. After tachyon condensation, the shift two-form $S$ translates into a flux on the D$7$-brane equal to 
$F_{{\rm D}7} = \tfrac{1}{2}\,D_P-S$. One can easily check that this charge vector indeed reproduces the right flux and curvature induced lower brane charges expected for a D$7$-brane \eqref{defsmoothd7charges}. 
The charge vector of the image brane is defined by using the image $\overline{{\rm D}9}$/D$9$ pair as $\Gamma_{{\rm D}7'} = \Gamma_1'+\Gamma_2$. One can easily check the the D$7$ charge will be $D_P$ again, and that the flux on the resulting D$7$ will be $-F_{{\rm D}7}+2\,B$.

For the sake of concreteness, let us work out the charge vector of the D$7_A$-brane of the first scenario of the first model in section \ref{sec:firstmodel}. In this case, the divisor and the desired worldvolume flux are
\begin{equation*}
D_P = \eta_4\,, \quad F_A = \tfrac{1}{2}\, \eta_4-S\,,  \qquad {\rm with} \quad S=-\sum_{i=1}^4 a_i\,\eta_i\,.
\end{equation*}
We can immediately write the charge vector as
\begin{eqnarray}
\Gamma &=& \Big( {\rm exp}(\eta_4-S)-{\rm exp}(-S)\Big)\, \Big(1+\frac{c_2(X)}{24}\Big) \nonumber\\
&=& \eta_4 + \eta_4\cdot (\tfrac{1}{2}\eta_4-S) + \Big(\tfrac{1}{2}\,\eta_4\,(\tfrac{1}{2}\,\eta_4-S)^2+\eta_4^3+c_2(X)\,\eta_4 \Big)\,,
\end{eqnarray}
where the last two terms give the Euler characteristic of the divisor, $\eta_4^3+c_2(X)\cdot \eta_4 = \chi(\eta_4)$. Hence, we see that it is clearly much more convenient to construct charge vectors by means of this method than by straightforward application of \eqref{defsmoothd7charges}.\\
As explained at the end of \ref{sec:dbranecharges&dsz}, this charge vector can only be used to compute induced charges, and deduce the net chiral spectrum of intersecting D7-branes. To find out about the non-chiral sector, more work is required.

\subsubsection{Whitney-type branes}
Let us now review how Whitney-type branes are treated in this language. In \cite{Collinucci:2008pf} the condition was derived that all orientifold-invariant configurations should actually be made out of an even number of D$9$/$\overline{{\rm D}9}$ pairs. In a sense, a Whitney-type brane can be thought of as a D$7$/image-D$7$ pair that has recombined into a single invariant brane. For a Whitney-type brane of even D$7$-charge $D_W=2\,D_P$, the charge vector is simply:
\begin{eqnarray}
\Gamma &=& \Gamma_1+\Gamma_2+\Gamma_1'+\Gamma_2'  \\
&=& \Big( {\rm ch}(D_P-S)+{\rm ch}(S+2\,B) -{\rm ch}(-D_P+S+2\,B) -{\rm ch}(-S)\Big)\,\Big( 1+ \frac{c_2(X)}{24} \Big)\,, \nonumber
\end{eqnarray}
where the $\Gamma$'s are the ones we defined before. One can easily check that this is involution-invariant, that the two-form component is indeed $D_W$, and that the four-form component is $D_W \cdot F_W = D_W \cdot B$, as expected. The choice of the two-form $S$ corresponds to adjusting the flux on the D$7_W$ of type $h^{1,1}_-(D_H)$. Define the involution as $\xi \rightarrow -\xi$. Then the D$7_W$-brane resulting from tachyon condensation will have a singular divisor equation given by 
\begin{equation}
\eta^2+\xi^2\,(\rho \, \tau-\psi^2)=0\,,
\end{equation}
 where $\{\eta\,; \psi\,; \rho\,; \tau\}$ are sections of the line bundles associated to the classes
 \begin{equation}
  \{D_P\,; D_P-D_{\xi}\,; 2\,(D_P-S-B)-D_{\xi}\,; 2\,(S+B)-D_{\xi}\}\,, 
  \end{equation}
respectively. In order for the D$7$-brane to retain its `structural integrity', one must choose $S$ such that all these bundles are positive definite, or else this will modify the structure of the brane severely. Starting with $\psi$, we see that as long as we do not choose to have a single D$7$/image-D$7$ pair on top of the O$7$-plane, this class will always be positive-definite. Should either one of the polynomials $\rho$ and $\tau$ correspond to a section of a negative bundle, which would not be globally well-defined, then we would have to set it identically to zero. In this case, the divisor equation would factorize into a D$7$/imag-D$7$ pair as follows:
 \begin{equation}
 \eta^2+\xi^2\,\psi^2 = 0 \quad \Rightarrow \quad \{\eta+\xi\,\psi = 0\}\, \cup \,\{\eta-\xi\,\psi = 0\}\,.
 \end{equation}
 The constraints for $\rho$ and $\tau$ to be globally well-defined are
 \begin{equation}
 D_P-\frac{[\xi]}{2}-B \geq S \geq \frac{[\xi]}{2}-B\,. \label{whitneyconstraints}
 \end{equation} 
 Fortunately, $S$ will drop out of the calculation of intersection products with the other present branes. It will, however, enter the D$3$ tadpole calculation.


\section{Third model} \label{sec:T1model}

\subsection{R2 resolution of the $\mathbb{P}^4_{1,1,2,2,6}\left( 12 \right) /\mathbb{Z}_2:1\,0\,1\,0\,0$
geometry}
We repeat the projective weights for this space for convenience in table \ref{tab:repeatforR2}.
\begin{table}[h!]
\begin{centering}
\begin{tabular}{|c|c|c|c|c|c|c|c||c|}
\hline 
$x_{1}$ & $x_{2}$ & $x_{3}$ & $x_{4}$ & $x_{5}$ & $x_{6}$ & $x_{7}$ & $x_{8}$ & \emph{p}\tabularnewline
\hline
\hline 
2 & 1 & 6 & 1 & 2 & 0 & 0 & 0 & 12\tabularnewline
\hline 
2 & 1 & 5 & 0 & 2 & 0 & 0 & 2 & 12\tabularnewline
\hline 
2 & 0 & 5 & 1 & 2 & 0 & 2 & 0 & 12\tabularnewline
\hline 
1 & 0 & 3 & 0 & 1 & 1 & 0 & 0 & 6\tabularnewline
\hline 
\end{tabular}
\par 
\end{centering}
\caption{Projective weights for the R1 resolution of $\mathbb{P}^4_{2,1,6,1,2}\left( 12 \right) /\mathbb{Z}_2:0\,0\,1\,1\,0$.}
\label{tab:repeatforR2}
\end{table} 

The Stanley-Reisner ideal reads
\begin{equation}
SR=\{x_{2}x_{3},x_{2}x_{4},x_{3}x_{4},x_{3}x_{6},x_{2}x_{8},x_{6}x_8,x_{1}x_{4}x_{5}x_7,x_1x_5x_6x_7,x_{1}x_{5}x_{7}x_8\}\,.
\end{equation}

The triple intersection numbers in the basis $\eta_{1}=D_{2}$, $\eta_{2}=D_{4}$,
$\eta_{3}=D_{6}$ and $\eta_{4}=D_{8}$ are encoded in

\begin{eqnarray*}
I_{3} & = & 9\eta_1^3+3\eta_2^3+8\eta_3^3-72\eta_4^3-3\eta_1^2\eta_3 \\
 &  & +3\eta_2^2\eta_3-12\eta_2^2\eta_4+\eta_1\eta_3^2-5\eta_2\eta_3^2+30\eta_2\eta_4^2 \,.
\end{eqnarray*}
The volumes of the corresponding divisors are
\begin{eqnarray*}
\tau_{1} & = & \frac{1}{2} \left(3 t_1-t_3\right)^2\,,\\
\tau_{2} & = & \frac{1}{2} \left(3 t_2^2 + 6 t_2 t_3 - 5 t_3^2 - 24 t_2 t_4 + 30 t_4^2\right)\,,\\
\tau_{3} & = & -\frac{1}{2} \left(t_1 + t_2 - 2 t_3\right) \left(3 t_1 - 3 t_2 + 4 t_3\right)\,,\\
\tau_{4} & = & -6 \left(t_2 - 3 t_4\right) \left(t_2 - 2 t_4\right)\,.
\end{eqnarray*}
The volume of the CY manifold is given by
\begin{eqnarray}
\mathcal{V} & =& \frac{1}{6} \left[-\frac{1}{15} \left(3 t_2 - 5 t_3\right)^3 - \frac{1}{3} \left(-3 t_1 + t_3\right)^3 - 
 \frac{3}{5} \left(5 t_4-2 t_2\right)^3 + 3 t_4^3\right]\\
 & = & \frac{\sqrt{2}}{3}\left[\frac{1}{2\sqrt{6}}\left(\tau_{1}+5\tau_{2}+3\tau_3+2\tau_4\right)^{\frac{3}{2}}-\frac{1}{10\sqrt{6}}\left(\tau_{1}+5\tau_{2}+3\tau_3\right)^{\frac{3}{2}}-\frac{1}{15}\left(\tau_1+3\tau_3\right)^{\frac{3}{2}}-\frac{1}{3}\tau_{1}^{\frac{3}{2}}\right]. \nonumber
\end{eqnarray} 
It has the expected Swiss cheese form. From this volume formula we
deduce the diagonal basis to be
\begin{eqnarray}
D_{a} & = &  \eta_{1}+5\eta_{2}+3\eta_3+2\eta_4\,,\nonumber\\
D_{b} & = & \eta_1+5\eta_2+3\eta_3\,, \nonumber \\
D_{c} & = & \eta_1+3\eta_3\,, \nonumber \\
D_{d} & = & \eta_1\,. \nonumber
\end{eqnarray} \label{diagt1model}
In this basis the total volume reads\[
\mathcal{V}=\frac{\sqrt{2}}{3}\left(\frac{1}{2\sqrt{6}}\tau_{a}^{\frac{3}{2}}-\frac{1}{10\sqrt{6}}\tau_{b}^{\frac{3}{2}}-\frac{1}{15}\tau_{c}^{\frac{3}{2}}-\frac{1}{3}\tau_{d}^{\frac{3}{2}}\right),\]
and the triple intersections can be rewritten as\[
I_{3}=24\,D_{a}^{3}+600\,D_{b}^{3}+225\,D_{c}^{3}+9\,D_{d}^{3}.\]\,

The K\"ahler cone conditions read as follows: 
\begin{eqnarray*}
t_2 - 2 t_4 &> & 0\,,\\
-t_2 + t_3 + t_4 &> & 0\,,\\
t_1 + t_2 - 2 t_3 & > & 0\,,\\
-3 t_1 + t_3 & > & 0\,.
\end{eqnarray*}

Searching for smooth, `small' cycles with holomorphic Euler characteristic equal to one, we find the following three surfaces
\begin{eqnarray*}
\{D_2,D_4,D_6\} &=& \{\eta_1, \eta_2, \eta_3\}\,,\\
{\rm with} \quad h^{1,1} &=& \{1\,, 7\,, 2  \}\,.
\end{eqnarray*}
The first surface is a $\mathbb{CP}^2$. By inspecting the intersection numbers, we see that the other two surfaces fail to be Del Pezzos, even though their Hodge numbers are consistent with those of  $dP_6,$ and the Hirzebruch surfaces $\mathbb{F}_n$ (for arbitrary $n$), respectively. Let us work out the topology of $D_4$ in more detail.\footnote{We are very grateful to H. Skarke for sharing this calculation with us.} The SR ideal shows that $x_2$ and $x_3$ cannot vanish on the surface. Hence, we gauge-fix them to one. Now the CY polynomial takes the following form:
\begin{equation}
P^{(6)}(x_1; x_5; x_6\,x_7)+x_7\,x_8 = 0\,,
\end{equation}
where the first term is some polynomial of degree six in the three arguments given. We can now define a map from this surface onto $\mathbb{CP}^2$ as follows:
\begin{equation}
(x_1 : x_5 : x_6 : x_7 : x_8) \mapsto (y_1: y_2: y_3) = (x_1 : x_5 : x_6\,x_7)\,.
\end{equation}
This map is a blow-down of our surface onto the projective plane. Now, we can distinguish two cases:
\begin{enumerate}
\item $x_7 \neq 0$. In this case, we gauge-fix $x_7=1$. Now we see that choosing a point on the $\mathbb{CP}^2$, which amounts to choosing $x_1, x_2$, and $x_6$, completely determines $x_8$, since it appears linearly in the CY equation.

\item $x_7=0$. In this case, the CY equation takes the form $P^{(6)}(x_1; x_5)=0$, and $x_6$ and $x_8$ are undetermined. This means, that the preimages of the six points on the $\mathbb{CP}^2$ with $(x_1 : x_5 : 0)$, where the $P^{(6)}(x_1; x_5)=0$ is satisfied are curves parametrized by $(x_6: x_8)$.
\end{enumerate}
Therefore, our surface is indeed the blow-up of the projective plane at six points. However, all six points lie on the line (the $\mathbb{CP}^1$) defined by $y_3=0$. Hence, they are not in generic positions, which is a requirement in order to have a $dP_6$.

\subsection{Scenarios in the third model}

By inspecting the intersection numbers of this CY we see that $\eta_1$ and $\eta_2$ do not intersect. Therefore, we only have two possible scenarios.
We summarize our results in table \ref{tab:scenariost1model}:

\begin{table}[h!]
\begin{centering}
\begin{tabular}{|c|c|c|c|}
\hline 
Scenario&E3 & D7$_A$ & D7$_B$
\tabularnewline
\hline
\hline 
I&$\eta_1$ & $\eta_2$ & $\eta_3$ \tabularnewline
\hline 
&arbitrary & arbitrary  & $\{1+e_1+n\,; b_2\,; 1+e_3+3\,n\,;b_4 \}$ \tabularnewline
\hline 
\hline
\hline
II&$\eta_2$ & $\eta_1$ & $\eta_3$ \tabularnewline
\hline 
&arbitrary & arbitrary & $\{b_1\,; 3+e_2+5\,n\,;1+e_3+3\,n\,;b_4 \}$ \tabularnewline
\hline
\end{tabular}
\par\end{centering}

\caption{Two `local' models.}
\label{tab:scenariost1model}
\end{table} 

For the global model, we will pick the involution $x_3 \rightarrow -x_3$. Solving the equations
\begin{equation}
\langle \Gamma_W\,, \, \Gamma_A \rangle = \langle \Gamma_W\,, \, \Gamma_B \rangle=0\,,
\end{equation}
we find the following solutions:
\begin{enumerate}
\item {\bf Scenario $I$}:
The constraints we get from setting the chiral intersections with the hidden brane to zero are the following:
\begin{align}
N_A &= 3\,N\,, \quad &N_B =& 5\,N\,, \quad &{\rm for \ some }& \quad N\in \mathbb{Z}\,,\\
a_2 &= 2+e_2+5\,t\,, \quad &a_4 =& 1+e_4+2\,t\,, \quad &{\rm for \ some}& \quad t \in \mathbb{Z}\,.
\end{align}
As we see here, this setup requires that we put further constraints on the `local' model. To get an idea of how much D3 tadpole this Whitney-type brane can induce, let us compute it for the `minimal' choice of the shift vector $S$ in formula \eqref{whitneyconstraints}:
\begin{equation}
Q_{W, D3} = 372-\frac{3}{2}\,N-197\,N^3\,.
\end{equation}
Finally, let us compute the FI-terms for both MSSM branes in light of these constraints:
\begin{equation}
\xi_A,\, \xi_B \propto \sqrt{\tau_c}\,,
\end{equation}
where $D_c = \eta_1+3\,\eta_3$.

The self intersection volume of the instanton in this scenario is given by
\begin{equation}
{\rm Vol}\left(D_{E3}\cap D_{E3}\right) = 9 t_1 -3 t_3 =-3\sqrt{2}\sqrt{\tau_b}\,.
\end{equation}

Looking at the K\"ahler cone in the diagonal basis
\begin{eqnarray*}
\sqrt{\tau_a}-5\sqrt{\tau_b} & > & 0\,,\\
2\sqrt{\tau_b}-3\sqrt{\tau_c} & > & 0\,,\\
5\sqrt{\tau_c}-\sqrt{\tau_d} & > & 0\,,\\
3\sqrt{\tau_d} & > & 0\, ,
\end{eqnarray*}
the third equation indicates that the volume of the instanton has to go to zero. The reason is that the D-term potential dominates in the LVS, and setting this term to zero means $\tau_c$ has to vanish. Having a volumeless instanton now ruins the LVS. Again, we expect this D-term to be corrected by string loops, which could salvage this LVS.

\item {\bf Scenario $II$}:\\
The only constraint we get from setting the chiral intersections to zero is $N_B = 3\,N_A$.
Let us also compute the D3 tadpole for this hidden brane with the `minimal' choice of $S$:
\begin{equation}
Q_{W, D3} = 372+\frac{3}{2}\,N_A-75\,N_A^3\,.
\end{equation}
In this case, both branes give again similar FI terms
\begin{equation}
\xi_A, \, \xi_B \propto \sqrt{\tau_1}\,.
\end{equation}
\end{enumerate}

The self intersection volume of the instanton in this scenario is given by
\begin{eqnarray}
{\rm Vol}\left(D_{E3}\cap D_{E3}\right) & = & 3 t_2 + 3 t_3 - 12 t_4 =-4\sqrt{3\tau_a} +\frac{1}{2}\sqrt{3\tau_b}-\frac{21}{2\sqrt{2}}\sqrt{\tau_c}\nonumber\\
& = & -\left( 4\sqrt{3\tau_a} -\frac{1}{2}\sqrt{3(\tau_c+5\tau_{E3})}+\frac{21}{2\sqrt{2}}\sqrt{\tau_c}\right)\,.
\end{eqnarray}

In this case, the same problem as in \eqref{equ:prob-f-term} occurs, by making the volume large we get an imaginary part in the solution for the volume. Thus, in the second scenario we do not get a large volume compactification either.


\section{Fourth model: A matterless model} \label{sec:nomatter}
The following model, as it turns out, will yield a trivial field content. Nevertheless, we will present the geometry in case the reader wants to use it differently.

\subsection{The resolved $\mathbb{P}^4_{1,1,1,3,3}\left(9\right)/\mathbb{Z}_3: 0\,0\,0\,2\,1$ geometry}

\begin{table}[h]
\begin{centering}
\begin{tabular}{|c|c|c|c|c|c|c|c||c|}
\hline 
$x_{1}$ & $x_{2}$ & $x_{3}$ & $x_{4}$ & $x_{5}$ & $x_{6}$ & $x_{7}$ & $x_{8}$ & \emph{p}\tabularnewline
\hline
\hline 
1 & 1 & 3 & 1 & 3 & 0 & 0 & 0 & 9\tabularnewline
\hline 
2 & 2 & 3 & 2 & 0 & 0 & 0 & 9 & 18\tabularnewline
\hline 
0 & 0 & 0 & 0 & 1 & 2 & 3 & 0 & 6\tabularnewline
\hline 
0 & 0 & 0 & 0 & 0 & 1 & 1 & 1 & 3\tabularnewline
\hline
\end{tabular}
\par
\end{centering}
\caption{Projective weights for the resolved $\mathbb{P}^4_{1,1,3,1,3}\left(9\right)/\mathbb{Z}_3: 0\,2\,1\,0\,0$ space.}
\end{table}\label{tab:charges-9-1-1-1-3-3}
The Stanley-Reisner ideal reads
\begin{equation}
SR=\{x_3x_5, x_5x_7,x_6x_7,x_3x_8, x_6x_8, x_1x_2x_4\}\,.
\end{equation}
The triple intersection numbers in the basis $\eta_1 = D_8$ , $\eta_2 = D_6$ , $\eta_3 = D_5$ , $\eta_4 = D_{1,2,4}$ are
encoded in
\begin{eqnarray*}
I_3 & = & -216\eta_1^3+9\eta_2^3+9\eta_3^3+\eta_3\eta_4^2+\eta_2\eta_4^2-3\eta_3^2\eta_4\\
& & -27\eta_1\eta_3^2 - 3\eta_2^2\eta_4-18\eta_1^2\eta_4+81\eta_1^2\eta_3+9\eta_1\eta_3\eta_4 \,.
\end{eqnarray*}

The volumes of the corresponding divisors are

\begin{eqnarray*}
\tau_{1} & = & -\frac{9}{2}\left(2t_{1}-t_{3}\right)\left(12t_{1}-3t_{3}+2t_{4}\right)\,,\\
\tau_{2} & = & \frac{1}{2}\left(3t_{2}-t_{4}\right)^{2}\,,\\
\tau_{3} & = & \frac{1}{2}\left(9t_{1}-3t_{3}+t_{4}\right)^{2}\,,\\
\tau_{4} & = & \frac{1}{2} \left(-18 t_1^2 - 3 t_2^2 + 18 t_1 t_3 - 3 t_3^2 + 2 t_2 t_4 + 2 t_3 t_4\right)\,.
\end{eqnarray*}

The volume of the Calabi-Yau manifold is given by
\begin{eqnarray*}
\mathcal{V} & = & \frac{1}{18}\left[3\left(3t_{1}+t_{4}\right)^{3}-t_{4}^{3}-\left(3t_{2}-t_{4}\right)^{3}-\left(9t_{1}-3t_{3}+t_{4}\right)^{3}\right]\\
& = &\frac{\sqrt{2}}{9} \left[\frac{1}{\sqrt{3}}\left(\tau_{1}+3\tau_{3}\right)^{\frac{3}{2}}-\left(\tau_{1}-\tau_{2}+2\tau_{3}-3\tau_{4}\right)^{\frac{3}{2}}-\tau_{2}^{\frac{3}{2}}-\tau_{3}^{\frac{3}{2}}\right].
\end{eqnarray*}
It has the expected Swiss cheese form. From this volume formula we
deduce the diagonal basis to be
\begin{eqnarray}
D_{a} & = & \eta_{1}+3\eta_{3}\,,\nonumber\\
D_{b} & = & \eta_{1}-\eta_{2}+2\eta_{3}-3\eta_{4}\,,\nonumber\\
D_{c} & = & \eta_{2}\,,\\ \label{equ:dia-div-9}
D_{d} & = & \eta_{3}\,, \nonumber
\end{eqnarray}
in this basis the total volume reads\[
\mathcal{V}=\frac{\sqrt{2}}{9}\left[\frac{1}{\sqrt{3}}\tau_{a}^{\frac{3}{2}}-\tau_{b}^{\frac{3}{2}}-\tau_{c}^{\frac{3}{2}}-\tau_{d}^{\frac{3}{2}}\right],\]
and the triple intersections can be rewritten as\[
I_{3}=27D_{a}^{3}+9D_{b}^{3}+9D_{c}^{3}+9D_{d}^{3}.\]

The K\"ahler cone conditions read as follows: 
\begin{eqnarray*}
-2 t_1 + t_3 &> & 0\,,\\
t_1 & > & 0\,,\\
3 t_1 + t_2 - t_3 & > & 0\,,\\
-3 t_2 + t_4 & > & 0\,,\\
t_2 & > & 0\,.
\end{eqnarray*}

Searching for rigid divisors with holomorphic Euler characteristic one, we find the following three solutions:
\begin{eqnarray*}
\{D_3,D_5,D_6\} &=& \{-3\,\eta_4+2\,\eta_3 -\eta_2+\eta_1\,, \eta_3\,, \eta_2\}\,,\\
{\rm with} \quad h^{1,1} &=& \{1\,, 1\,, 1 \}\,.
\end{eqnarray*}
Hence, all three are $\mathbb{CP}^2$'s. The striking feature about these divisors, which ultimately kills the model for our purposes, lies in the fact that no two of them intersect. Although this automatically solves the problem of unwanted zero modes, it does so too drastically, as no chiral matter can arise from D7-branes wrapped on them.

Inspecting the available involutions, we see that it is impossible to have two D7-branes on distinct cycles and cancel the D7 tadpole. Hence, one can only have a single D7-brane, and in this case, it must be on top of the O7-plane.

\subsection{Moduli stabilization}
Although we can not do any model building in this example we can nevertheless look at the stabilization problem.
So first we choose divisors on which we would like to wrap our D7-branes. However by inspecting table \ref{tab:charges-9-1-1-1-3-3} carefully we see that we can have only one brane if we want to compensate D7-charge only via an orientifold plan. Hence we will work with a configuration where we have a D7-brane on a divisor four times the divisor class of the orientifold and nothing else. If we wrap the brane on a diagonal divisor we obtain that the FI term is proportional to its volume. Knowing this we have to take a divisor that is unrestrictedly shrinkable. For this we rewrite the K\"ahler cone in terms of the diagonal basis
\begin{eqnarray*}
\sqrt{\tau_a}-\sqrt{\tau_c} &> & 0\,,\\
\sqrt{\tau_a}-\sqrt{\tau_b} & > & 0\,,\\
\sqrt{\tau_c}-\sqrt{\tau_d}& > & 0\,,\\
3\sqrt{\tau_d}& > & 0\,,\\
\sqrt{\tau_b}-\sqrt{\tau_d}& > & 0\,.
\end{eqnarray*}
Having the D7-brane on $D_d$ we can put the instanton either on $D_b$ or $D_c$. For these two cases we get the following selfintersection volumes for the instanton
\begin{equation}
{\rm Vol}\left(D_b\cap D_b\right) = -3 t_4=-3 \sqrt{2 \tau_b}\,,
\end{equation}
and
\begin{equation}
{\rm Vol}\left(D_c\cap D_c\right) = -3 t_4 + 9 t_1 - 27 t_3=-3 \sqrt{2 \tau_c}\, ,
\end{equation}
respectively. In this case, the potentials for the two scenarios are symmetric under exchange of $\tau_b$ and $\tau_c$. With \eqref{equ:f-term} and $A_{E3}=1$, $|W_0|=5$ and $g_s=\frac{1}{10}$ we find for both minima
\begin{equation}
 \tau_{E3}= 1.41\,, \quad {\cal V} = 6.74\cdot 10^{36}\,,
\end{equation}
where in each case one flat direction remains. Therefore, in this trivial model we can only stabilize three out of the four moduli.

\bibliographystyle{JHEP}
\bibliography{articles2}

\providecommand{\href}[2]{#2}\begingroup\raggedright\begin{thebibliography}{10}

\bibitem{Balasubramanian:2005zx}
V.~Balasubramanian, P.~Berglund, J.~P. Conlon, and F.~Quevedo, {\it
  {Systematics of moduli stabilisation in Calabi-Yau flux compactifications}},
  {\em JHEP} {\bf 03} (2005) 007,
  [\href{http://arxiv.org/abs/hep-th/0502058}{{\tt hep-th/0502058}}].

\bibitem{Kachru:2003aw}
S.~Kachru, R.~Kallosh, A.~Linde, and S.~P. Trivedi, {\it {De Sitter vacua in
  string theory}},  {\em Phys. Rev.} {\bf D68} (2003) 046005,
  [\href{http://arxiv.org/abs/hep-th/0301240}{{\tt hep-th/0301240}}].

\bibitem{Denef:2008wq}
F.~Denef, {\it {Les Houches Lectures on Constructing String Vacua}},
  \href{http://arxiv.org/abs/0803.1194}{{\tt arXiv:0803.1194}}.

\bibitem{Wijnholt:2008db}
M.~Wijnholt, {\it {F-Theory, GUTs and Chiral Matter}},
  \href{http://arxiv.org/abs/0809.3878}{{\tt arXiv:0809.3878}}.

\bibitem{Blumenhagen:2007sm}
R.~Blumenhagen, S.~Moster, and E.~Plauschinn, {\it {Moduli Stabilisation versus
  Chirality for MSSM like Type IIB Orientifolds}},  {\em JHEP} {\bf 01} (2008)
  058, [\href{http://arxiv.org/abs/0711.3389}{{\tt arXiv:0711.3389}}].

\bibitem{Freed:1999vc}
D.~S. Freed and E.~Witten, {\it {Anomalies in string theory with D-branes}},
  \href{http://arxiv.org/abs/hep-th/9907189}{{\tt hep-th/9907189}}.

\bibitem{Kreuzer:2000xy}
M.~Kreuzer and H.~Skarke, {\it {Complete classification of reflexive polyhedra
  in four dimensions}},  {\em Adv. Theor. Math. Phys.} {\bf 4} (2002)
  1209--1230, [\href{http://arxiv.org/abs/hep-th/0002240}{{\tt
  hep-th/0002240}}].

\bibitem{Kreuzer:2002uu}
M.~Kreuzer and H.~Skarke, {\it {PALP: A Package for analyzing lattice polytopes
  with applications to toric geometry}},  {\em Comput. Phys. Commun.} {\bf 157}
  (2004) 87--106, [\href{http://arxiv.org/abs/math/0204356}{{\tt
  math/0204356}}].

\bibitem{wwwCY}
 \href{http://hep.itp.tuwien.ac.at/~kreuzer/CY}{http://hep.itp.tuwien.ac.at/~k%
reuzer/CY}.

\bibitem{SINGULAR}
G.-M. Greuel, G.~Pfister, and H.~Sch{\"o}nemann, {\it {Singular 3.0.2. A
  Computer Algebra System for Polynomial Computations. Centre for Computer
  Algebra}},  {\em University of Kaiserslautern} (2006).
  \href{http://www.singular.uni-kl.de} {http://www.singular.uni-kl.de}.

\bibitem{Cicoli:2008va}
M.~Cicoli, J.~P. Conlon, and F.~Quevedo, {\it {General Analysis of LARGE Volume
  Scenarios with String Loop Moduli Stabilisation}},  {\em JHEP} {\bf 10}
  (2008) 105, [\href{http://arxiv.org/abs/0805.1029}{{\tt arXiv:0805.1029}}].

\bibitem{Gukov:1999ya}
S.~Gukov, C.~Vafa, and E.~Witten, {\it {CFT's from Calabi-Yau four-folds}},
  {\em Nucl. Phys.} {\bf B584} (2000) 69--108,
  [\href{http://arxiv.org/abs/hep-th/9906070}{{\tt hep-th/9906070}}].

\bibitem{Becker:2002nn}
K.~Becker, M.~Becker, M.~Haack, and J.~Louis, {\it {Supersymmetry breaking and
  alpha'-corrections to flux induced potentials}},  {\em JHEP} {\bf 06} (2002)
  060, [\href{http://arxiv.org/abs/hep-th/0204254}{{\tt hep-th/0204254}}].

\bibitem{Bobkov:2004cy}
K.~Bobkov, {\it {Volume stabilization via alpha' corrections in type IIB theory
  with fluxes}},  {\em JHEP} {\bf 05} (2005) 010,
  [\href{http://arxiv.org/abs/hep-th/0412239}{{\tt hep-th/0412239}}].

\bibitem{Cicoli:2007xp}
M.~Cicoli, J.~P. Conlon, and F.~Quevedo, {\it {Systematics of String Loop
  Corrections in Type IIB Calabi- Yau Flux Compactifications}},  {\em JHEP}
  {\bf 01} (2008) 052, [\href{http://arxiv.org/abs/0708.1873}{{\tt
  arXiv:0708.1873}}].

\bibitem{Evslin:2006cj}
J.~Evslin, {\it {What does(n't) K-theory classify?}},
  \href{http://arxiv.org/abs/hep-th/0610328}{{\tt hep-th/0610328}}.

\bibitem{Minasian:1997mm}
R.~Minasian and G.~W. Moore, {\it {K-theory and Ramond-Ramond charge}},  {\em
  JHEP} {\bf 11} (1997) 002, [\href{http://arxiv.org/abs/hep-th/9710230}{{\tt
  hep-th/9710230}}].

\bibitem{Denef:2007vg}
F.~Denef and G.~W. Moore, {\it {Split states, entropy enigmas, holes and
  halos}},  \href{http://arxiv.org/abs/hep-th/0702146}{{\tt hep-th/0702146}}.

\bibitem{Akerblom:2007nh}
N.~Akerblom, R.~Blumenhagen, D.~Lust, and M.~Schmidt-Sommerfeld, {\it {D-brane
  Instantons in 4D Supersymmetric String Vacua}},  {\em Fortsch. Phys.} {\bf
  56} (2008) 313--323, [\href{http://arxiv.org/abs/0712.1793}{{\tt
  arXiv:0712.1793}}].

\bibitem{Blumenhagen:2006xt}
R.~Blumenhagen, M.~Cvetic, and T.~Weigand, {\it {Spacetime instanton
  corrections in 4D string vacua - the seesaw mechanism for D-brane models}},
  {\em Nucl. Phys.} {\bf B771} (2007) 113--142,
  [\href{http://arxiv.org/abs/hep-th/0609191}{{\tt hep-th/0609191}}].

\bibitem{akerblomphd}
N.~Akerblom, {\it {D-instantons and Effective Couplings in Intersecting D-brane
  Models}},  {\em Ph.D. Thesis} (2008).

\bibitem{Kachru:2008wt}
S.~Kachru and D.~Simic, {\it {Stringy Instantons in IIB Brane Systems}},
  \href{http://arxiv.org/abs/0803.2514}{{\tt arXiv:0803.2514}}.

\bibitem{Argurio:2007qk}
R.~Argurio, M.~Bertolini, S.~Franco, and S.~Kachru, {\it {Metastable vacua and
  D-branes at the conifold}},  {\em JHEP} {\bf 06} (2007) 017,
  [\href{http://arxiv.org/abs/hep-th/0703236}{{\tt hep-th/0703236}}].

\bibitem{Argurio:2007vqa}
R.~Argurio, M.~Bertolini, G.~Ferretti, A.~Lerda, and C.~Petersson, {\it
  {Stringy Instantons at Orbifold Singularities}},  {\em JHEP} {\bf 06} (2007)
  067, [\href{http://arxiv.org/abs/0704.0262}{{\tt arXiv:0704.0262}}].

\bibitem{Bianchi:2007wy}
M.~Bianchi, F.~Fucito, and J.~F. Morales, {\it {D-brane Instantons on the
  $T^6/Z_3$ orientifold}},  {\em JHEP} {\bf 07} (2007) 038,
  [\href{http://arxiv.org/abs/0704.0784}{{\tt arXiv:0704.0784}}].

\bibitem{Ibanez:2007rs}
L.~E. Ibanez, A.~N. Schellekens, and A.~M. Uranga, {\it {Instanton Induced
  Neutrino Majorana Masses in CFT Orientifolds with MSSM-like spectra}},  {\em
  JHEP} {\bf 06} (2007) 011, [\href{http://arxiv.org/abs/0704.1079}{{\tt
  arXiv:0704.1079}}].

\bibitem{Petersson:2007sc}
C.~Petersson, {\it {Superpotentials From Stringy Instantons Without
  Orientifolds}},  {\em JHEP} {\bf 05} (2008) 078,
  [\href{http://arxiv.org/abs/0711.1837}{{\tt arXiv:0711.1837}}].

\bibitem{Kreuzer:2006ax}
M.~Kreuzer, {\it {Toric Geometry and Calabi-Yau Compactifications}},
  \href{http://arxiv.org/abs/hep-th/0612307}{{\tt hep-th/0612307}}.

\bibitem{GH}
P.~Griffiths and J.~Harris, {\it {Principles of Algebraich Geometry}},  {\em
  Wiley} (1978).

\bibitem{GarciaEtxebarria:2007zv}
I.~Garcia-Etxebarria and A.~M. Uranga, {\it {Non-perturbative superpotentials
  across lines of marginal stability}},  {\em JHEP} {\bf 01} (2008) 033,
  [\href{http://arxiv.org/abs/0711.1430}{{\tt arXiv:0711.1430}}].

\bibitem{GarciaEtxebarria:2008pi}
I.~Garcia-Etxebarria, F.~Marchesano, and A.~M. Uranga, {\it {Non-perturbative
  F-terms across lines of BPS stability}},  {\em JHEP} {\bf 07} (2008) 028,
  [\href{http://arxiv.org/abs/0805.0713}{{\tt arXiv:0805.0713}}].

\bibitem{Beasley:2004ys}
C.~Beasley and E.~Witten, {\it {New instanton effects in supersymmetric QCD}},
  {\em JHEP} {\bf 01} (2005) 056,
  [\href{http://arxiv.org/abs/hep-th/0409149}{{\tt hep-th/0409149}}].

\bibitem{Beasley:2005iu}
C.~Beasley and E.~Witten, {\it {New instanton effects in string theory}},  {\em
  JHEP} {\bf 02} (2006) 060, [\href{http://arxiv.org/abs/hep-th/0512039}{{\tt
  hep-th/0512039}}].

\bibitem{Collinucci:2008pf}
A.~Collinucci, F.~Denef, and M.~Esole, {\it {D-brane Deconstructions in IIB
  Orientifolds}},  \href{http://arxiv.org/abs/0805.1573}{{\tt
  arXiv:0805.1573}}.

\bibitem{Blumenhagen:2008at}
R.~Blumenhagen, V.~Braun, T.~W. Grimm, and T.~Weigand, {\it {GUTs in Type IIB
  Orientifold Compactifications}},  \href{http://arxiv.org/abs/0811.2936}{{\tt
  arXiv:0811.2936}}.

\bibitem{Grimm:2008ed}
T.~W. Grimm and A.~Klemm, {\it {U(1) Mediation of Flux Supersymmetry
  Breaking}},  {\em JHEP} {\bf 10} (2008) 077,
  [\href{http://arxiv.org/abs/0805.3361}{{\tt arXiv:0805.3361}}].

\bibitem{Brunner:2003zm}
I.~Brunner and K.~Hori, {\it {Orientifolds and mirror symmetry}},  {\em JHEP}
  {\bf 11} (2004) 005, [\href{http://arxiv.org/abs/hep-th/0303135}{{\tt
  hep-th/0303135}}].

\bibitem{Gimon:1996rq}
E.~G. Gimon and J.~Polchinski, {\it {Consistency Conditions for Orientifolds
  and D-Manifolds}},  {\em Phys. Rev.} {\bf D54} (1996) 1667--1676,
  [\href{http://arxiv.org/abs/hep-th/9601038}{{\tt hep-th/9601038}}].

\end{thebibliography}\endgroup

\end{document}